\documentclass[journal,twoside,web]{ieeecolor}
\usepackage{tmi}
\usepackage{cite}
\usepackage{amsmath,amssymb,amsfonts}
\usepackage{graphicx}
\usepackage{textcomp}
\usepackage{hyperref}

\usepackage{booktabs,multirow}
\usepackage{algorithm}
\usepackage{algpseudocode}
\usepackage{pifont}
\usepackage{xcolor}  

\usepackage{etoolbox}
\makeatletter
\@ifundefined{color@begingroup}%
{\let\color@begingroup\relax
\let\color@endgroup\relax}{}%
\def\fix@ieeecolor@hbox#1{%
\hbox{\color@begingroup#1\color@endgroup}}
\patchcmd\@makecaption{\hbox}{\fix@ieeecolor@hbox}{}{FAILED}
\patchcmd\@makecaption{\hbox}{\fix@ieeecolor@hbox}{}{FAILED}
\makeatother

\newcommand{\red}[1]{\textcolor{red}{#1}}
\colorlet{red}{black}

\def\BibTeX{{\rm B\kern-.05em{\sc i\kern-.025em b}\kern-.08em
    T\kern-.1667em\lower.7ex\hbox{E}\kern-.125emX}}
\usepackage{colortbl} 
\usepackage{nicematrix} 

\definecolor{posblue}{RGB}{0,0,200}
\definecolor{negred}{RGB}{200,0,0}
\definecolor{ourcolor}{gray}{0.95}

\begin{document}
\title{Cross-Distribution Diffusion Priors-Driven Iterative Reconstruction for Sparse-View CT}
\author{Haodong Li, Shuo Han, Haiyang Mao, Yu Shi, Changsheng Fang, Jianjia Zhang, Weiwen Wu, \IEEEmembership{Member, IEEE}, Hengyong Yu, \IEEEmembership{Fellow, IEEE}
\thanks{ This work was supported in part by NIH/NIBIB under grants R01EB034737 and R01EB032807.}
\thanks{ H. Li, S. Han, Y. Shi, C. Fang and H. Yu are with the Department of Electrical and Computer Engineering, University of Massachusetts Lowell, Lowell, MA, USA. H. Mao, J. Zhang and W. Wu are with the School of Biomedical Engineering, Sun Yat-Sen University, Shenzhen, Guangdong, China. Correspondence should be addressed to W. Wu (Email: wuweiw7@mail.sysu.edu.cn) or  H. Yu (Email:hengyong-yu@ieee.org).}
}

\maketitle


\begin{abstract}
Sparse-View CT (SVCT) reconstruction improves temporal resolution and reduces radiation dose, yet its clinical use is hindered by artifacts due to view reduction and domain shifts from scanner, protocol, or anatomical variations, leading to performance degradation in out-of-distribution (OOD) scenarios. 
We propose a Cross-Distribution Diffusion Priors-Driven Iterative Reconstruction (CDPIR) framework to tackle the OOD problem in SVCT. CDPIR integrates cross-distribution diffusion priors, derived from a Scalable Interpolant Transformer (SiT), with model-based iterative reconstruction methods. Specifically, we train a SiT backbone, an extension of the Diffusion Transformer (DiT) architecture, to establish a unified stochastic interpolant framework, leveraging Classifier-Free Guidance (CFG) across multiple datasets. 
\red{By randomly dropping the conditioning with a null embedding during training, the model learns a more transferable cross-distribution prior that encourages domain-invariant anatomical structures while allowing domain-specific appearance modulation.} During sampling, the globally sensitive transformer-based diffusion model exploits the cross-distribution prior within the unified stochastic interpolant framework, enabling flexible and stable control over multi-distribution-to-noise interpolation paths and decoupled sampling strategies, thereby improving adaptation to OOD reconstruction. By alternating between data fidelity and sampling updates, our model achieves state-of-the-art performance with superior detail preservation in SVCT reconstructions. Extensive experimental results demonstrate that CDPIR significantly outperforms existing approaches, particularly under OOD conditions, highlighting its robustness and potential clinical value in challenging imaging scenarios. 
The code is available at \url{https://github.com/Graeme-Lee/CDPIR}.

\end{abstract}


\section{Introduction}
\label{sec:introduction}
Sparse-view computed tomography (SVCT) has emerged as an effective strategy to reduce radiation dose and accelerate image acquisition. By minimizing the number of required angular projections, SVCT improves the efficiency of advanced CT systems, particularly in time-sensitive scenarios. The benefits of SVCT are especially pronounced in Medipix detector-based photon-counting CT (PCCT), also known as spectral CT, for extremity imaging \cite{willemink2018photon,walsh2011first, aamir2014mars, li2024deep}, which enables functional and molecular imaging with novel contrast agents. Compared with conventional CT, PCCT acquires rich spectral information but typically suffers from long scan durations and substantial radiation exposure, especially in extremity applications where high spatial resolution is essential. SVCT directly alleviates these limitations by shortening acquisition time and reducing dose, making spectral imaging more practical for longitudinal studies, dynamic contrast-enhanced protocols, and motion-prone clinical settings. Nonetheless, SVCT introduces severe artifacts and the risk of losing fine structural details. Consequently, developing robust reconstruction algorithms is critical to fully realize the clinical potential of SVCT and to improve the practical utility of extremity PCCT systems.

In recent years, many methods have been developed to address SVCT image reconstruction issue. Traditional model-based iterative reconstruction (MBIR) techniques \cite{ tovsic2011dictionary, buades2011non} have been widely explored. They impose prior assumptions onto the solution space to reduce artifacts, but often struggle with preserving high-frequency details and require extensive tuning. Recently, deep learning has emerged as a powerful alternative. CNN-based methods (\emph{e.g.,} FBPConvNet \cite{jin2017deep} and RED-CNN \cite{chen2017low}) employ paired training data to learn image priors and directly map low-quality reconstructions to their high-quality counterparts. These methods have demonstrated remarkable performance improvements over traditional approaches, mainly due to their strong inductive biases and ability to learn hierarchical features. In addition, Transformer-based networks have recently shown promise in SVCT reconstruction by capturing long-range dependencies and global contexts \cite{xia2022transformer,pan2022multi,li2025td}. Although their high computational cost poses challenges for large CT images, efficient variants such as the Restoration Transformer (Restormer) \cite{Zamir2021Restormer} offer scalable designs to balance local detail preservation with global context modeling.

In practice, obtaining paired training targets can be difficult, especially for PCCT where matched reference data are often unavailable. This has motivated self-supervised and zero-shot strategies \cite{guo2023spectral2spectral}\cite{niu2022noise}\cite{shi2025zs4d} to reduce reliance on clean targets or retraining.
Despite their success, most CNN and Transformer methods and many self-supervised and zero-shot strategies remain \emph{discriminative point-estimators} trained with pixel-wise objectives (\emph{e.g.}, MSE), which tends to yield conditional-mean solutions and produces over-smoothed reconstructions with attenuated high-frequency textures. 
These limitations have motivated a shift toward \emph{generative} priors for reconstruction, particularly denoising diffusion probabilistic models (DDPMs) \cite{DDPM} and score-based generative models (SGMs) \cite{song2019generative,song2020score,zhang2025improving}, which can model complex image distributions and provide posterior-like sampling for ill-posed inverse problems. Diffusion-based inverse solvers (DIS), such as DiffusionMBIR \cite{chung2023solving} and the Decomposed Diffusion Sampler (DDS) \cite{chung2023decomposed}, further integrate measurement fidelity by enforcing data consistency during the generative trajectory, enabling iterative refinement guided by both physics and learned priors. Recent advances also make diffusion-based reconstruction increasingly practical via scalable backbones and reduced-step sampling schemes.

\begin{figure}
    \centering
    \includegraphics[width=0.48\textwidth]{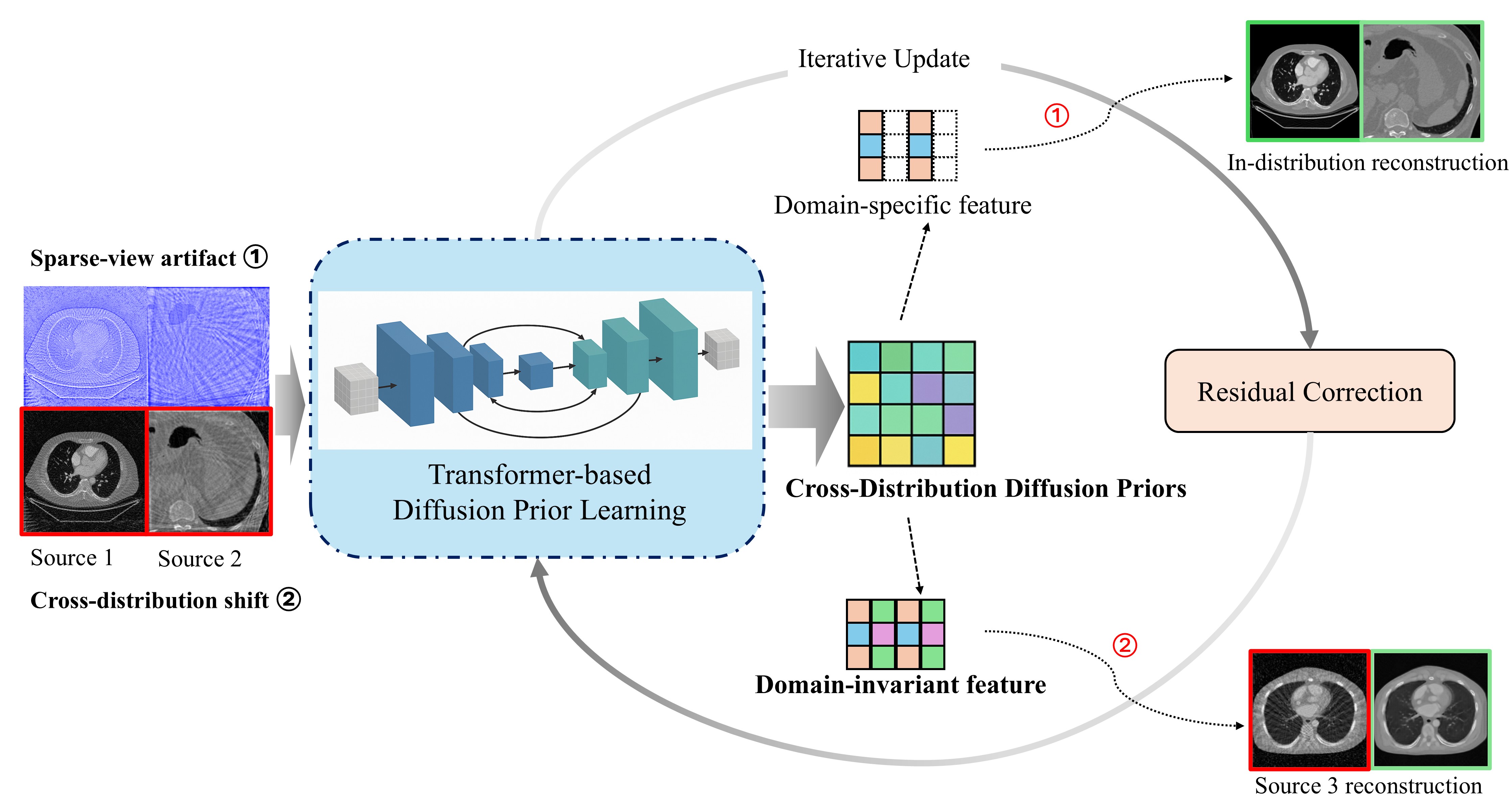}
    \caption{Overview of the proposed CDPIR framework. 
    Sparse-view CT inputs from multiple domains are processed by a diffusion transformer that disentangles domain-invariant features to mitigate distribution shifts and domain-specific features with residual correction to suppress artifacts, yielding robust reconstructions across heterogeneous clinical settings.
    Note that noisy images are highlighted with red boxes, and high-quality reconstructions are shown in green boxes.}
    \label{fig:OOD}
\end{figure}

In real-world deployment, SVCT reconstruction also faces substantial distribution shifts between training and testing data. For example, the AAPM Low-Dose CT Grand Challenge dataset~\cite{mccollough2017low} was acquired on Siemens SOMATOM scanners with abdominal protocols, whereas the Stanford COCA dataset~\cite{aimi2022coca} consists of coronary calcium and chest CT scans with different anatomical foci, acquisition protocols, and unspecified reconstruction parameters. Such cross-scanner and cross-protocol variations lead to out-of-distribution (OOD) scenarios in which testing measurements deviate from the training distribution, causing CNN-, Transformer-, and diffusion-based reconstructions to degrade markedly. These observations highlight two fundamental challenges (Fig.~\ref{fig:OOD}): 
\ding{172}~\textbf{Sparse-view artifacts}, \emph{i.e.}, severe streak artifacts and loss of structural fidelity arising from undersampled projections; and 
\ding{173}~\textbf{Cross-domain distribution differences}, \emph{i.e.}, scanner- and protocol-dependent variations that induce OOD scenarios.

Importantly, under the combined effects of ill-posed sparse- view acquisition and OOD shift, learning-based priors may exhibit a qualitatively different failure mode: hallucinations— visually plausible yet incorrect structures that violate data constraints. This is particularly concerning in medical imaging, as hallucinated textures or structures may mimic pathology or obscure subtle lesions, posing diagnostic risks.
Recent efforts to improve robustness under cross-domain shifts can be broadly grouped into three directions. 
(1) \emph{Test-time adaptation} adapts a pre-trained prior to the target measurements during reconstruction. For example, SCD injects lightweight LoRA modules and performs data-consistency-driven adaptation at sampling time to improve OOD performance, while DDIP3D extends this idea to efficient and coherent 3D adaptation for volumetric inverse problems~\cite{barbano2025steerable,chung2024deep}. 
(2) \emph{Large-capacity transformer models} mitigate OOD degradation by scaling representation power and long-range context modeling. For example, X-GRM employs a scalable transformer-based architecture and reports strong reconstruction quality on both in-domain and out-of-domain sparse-view inputs~\cite{liu2025x}. 
(3) \emph{Zero-shot score-based solvers} aim to handle distribution shifts without task-specific retraining by coupling a pretrained diffusion/score prior with explicit data fidelity. For example, Variational Score Solver (VSS) optimizes a variational objective that combines a diffusion prior term with a data-consistency term for SVCT reconstruction, enabling distribution-aware reconstruction and, in some settings, zero-shot deployment~\cite{he2024solving}.


Despite encouraging progress, these directions still have limitations for robust clinical applications: test-time adaptation can incur per-case optimization overhead; large-capacity models improve average robustness but do not fully control epistemic uncertainty under rare OOD cases; and diffusion priors may introduce \emph{stochastic hallucinations} if semantic priors dominate measurement constraints in severely ill-posed cases. These observations suggest that robust OOD SVCT reconstruction requires (i) a \emph{scalable, cross-distribution generative prior} to represent plausible anatomy across heterogeneous scanners and protocols, and (ii) an explicit \emph{physics-based iterative data-consistency mechanism} to enforce measured projections for previously unseen sinograms.

Motivated by this principle, we propose a \textbf{C}ross-distribution \textbf{D}iffusion \textbf{P}riors-driven \textbf{I}terative \textbf{R}econstruction (CDPIR) framework (Fig.~\ref{fig:OOD}). The main innovations of CDPIR focus on the prior side through two components. First, it employs the Scalable Interpolant Transformer (SiT)~\cite{ma2024sit} as the generative backbone, which supports systematic \emph{model-capacity scaling} and \emph{global context modeling} within the stochastic interpolant formulation. Second, it leverages classifier-free guidance (CFG)~\cite{ho2022classifierfreediffusionguidance} to enable \emph{cross-distribution training} via domain tokens. By randomly dropping the domain token during training, CDPIR jointly learns a domain-conditioned branch and a null-conditioned branch, encouraging the diffusion prior to capture anatomy-level structure shared across the training domains while modeling domain-dependent appearance statistics (\emph{e.g.}, texture, noise, and edge sharpness). Together, SiT and CFG yield a cross-distribution stochastic-interpolant diffusion prior that is both expressive and scalable.
Finally, CDPIR integrates this cross-distribution prior into a residual-guided iterative reconstruction pipeline with projection-domain data-consistency updates. This integration restricts generative refinement to the uncertainty permitted by the measured sinogram, thereby suppressing sparse-view artifacts while mitigating prior-driven hallucinations under cross-domain OOD shifts.

Our main contributions are summarized as follows:
\begin{itemize}
\item We propose CDPIR, a residual-guided conditional diffusion framework for SVCT reconstruction that integrates a transformer-based diffusion model with a physics-driven model-based iterative reconstruction pipeline. It is the first to unify conditional diffusion, stochastic interpolant sampling, and residual-guided updates within a single iterative reconstruction framework.

\item \red{We introduce a cross-distribution SVCT reconstruction framework that integrates classifier-free guidance (CFG) with a Transformer-based diffusion prior, encouraging a shared unconditional branch for broadly transferable anatomical structures and a token-conditioned residual for domain-specific appearance variations.} The resulting CDPIR achieves state-of-the-art OOD generalization and consistently outperforms representative diffusion- and CNN-based baselines (\emph{e.g.}, DDIP3D and DDS) on unseen distributions.

\item We extensively validate CDPIR on five real and synthetic datasets, covering different acquisition settings and artifact patterns. Our results demonstrate that it remains stable and accurate for a single CDPIR model trained across heterogeneous domains  under severe distribution shifts, highlighting its potential as a general-purpose medical image reconstruction framework.
\end{itemize}

\section{Method}
 The state-of-the-art (SOTA) methods (\emph{e.g.,} DDS) can infer clear images effectively when the testing data follow the same distribution as the training dataset. However, they face significant challenges for OOD reconstruction tasks. To address this issue, the proposed CDPIR employs a diffusion transformer framework combined with a residual-guided, projection-based projection constraint. This design effectively guides the transformer to reconstruct a high-quality CT image from the given projections.

\subsection{Sparse-view Iterative Reconstruction}
SVCT reconstruction is a linear inverse problem and can be mathematically formalized as:
\begin{equation}
\mathbf{y} = \mathbf{A} \mathbf{x} + \epsilon,
\end{equation}   
where $\mathbf{x} \in \mathbb{R}^n $ is vectorized CT image, $\mathbf{y}\in \mathbb{R}^m$ is the vectorized sinogram, $\mathbf{A}\in\mathbb{R}^{m \times n} $ is a system matrix, and $ \epsilon $ is the addictive noise in this process. $n$ and $m$ represent the dimensions of the image and sinogram, respectively.  
The objective is to reconstruct $\mathbf{x}$ from its measurement $\mathbf{y}$. Because typically $m<n$, the problem is ill-posed. In SVCT, the limited number of projections makes it even worse, resulting in artifacts, blurring, and noise.
To address this issue, traditional iterative reconstruction methods employ a residual-guided process. 
In this paper, we employ an improved Adaptive Steepest Descent Projection Onto Convex Sets (ASD-POCS) method with an improved total variation(iTV)\cite{ritschl2011improved} to guide the diffusion model to generate high-quality images.

\subsection{Scalable Interpolant Diffusion Transformer (SiT)}
SiT integrates a stochastic interpolant formulation into the Diffusion Transformer architecture~\cite{peebles2023scalable}, providing a unified view to support stochastic sampling through SDEs and deterministic sampling through ODEs~\cite{coddington1956theory}.
Instead of directly parameterizing the score function, SiT predicts a velocity field under a stochastic interpolant.
It defines an explicit interpolating process
\begin{equation}
\mathbf{x}_t=\alpha_t\mathbf{x}_0+\sigma_t\boldsymbol{\epsilon},\quad
\boldsymbol{\epsilon}\sim\mathcal{N}\!\left(\mathbf{0},\mathbf{I}\right),\quad t\in[0,1],
\label{eq:interpolant}
\end{equation}
where $\mathbf{x}_0$ is a clean data sample, $\boldsymbol{\epsilon}$ is stochastic noise, and $\alpha_t$ and $\sigma_t$ are time-dependent signal and noise coefficients.
Typically, $\alpha_0=1,\sigma_0=0$ and $\alpha_1=0,\sigma_1=1$.
In a VP-consistent setting, one may enforce $\alpha_t^2+\sigma_t^2=1$.
To make the interpolant fully specified and reproducible, we adopt a \textbf{linear interpolant}:
\begin{equation}
    \alpha_t = 1 - t, \quad \sigma_t = t, \quad t \in [0, 1].
\end{equation}

Under this framework, SiT samples the same marginals $p_t$ by integrating the reverse-time SDE
\begin{equation}
d\mathbf{x}_t
=
\mathbf{v}(\mathbf{x}_t,t)\,dt
+\frac{1}{2}\omega_t\,\mathbf{s}(\mathbf{x}_t,t)\,dt
+\sqrt{\omega_t}\,d\bar{\mathbf{W}}_t,
\label{eq:sit}
\end{equation}
where $\mathbf{s}(\mathbf{x},t)=\nabla_{\mathbf{x}}\log p_t(\mathbf{x})$ is the score field and
$\mathbf{v}(\mathbf{x},t)$ is a deterministic velocity field (probability flow ODE).
A key property of SiT is that the score can be recovered from the velocity field:

\begin{equation}
\mathbf{s}(\mathbf{x},t)
=
\frac{1}{\sigma_t}\,
\frac{\alpha_t\,\mathbf{v}(\mathbf{x},t)-\dot{\alpha}_t\,\mathbf{x}}
{\dot{\alpha}_t\sigma_t-\alpha_t\dot{\sigma}_t},
\label{eq:s_from_v}
\end{equation} 
where $\dot{\alpha}_t=d\alpha_t/dt$ and $\dot{\sigma}_t=d\sigma_t/dt$.
For the reverse-time sampling in Eq.~\eqref{eq:sit}, we set $\omega_t=\sigma_t=t$.
This choice keeps the SDE drift term $\omega_t\,\mathbf{s}(\mathbf{x},t)$ bounded as $t\to 0$
(by canceling the $1/\sigma_t$ factor), improving numerical stability in the final refinement steps.
The derivation of Eq.~\eqref{eq:s_from_v} can be found in~\cite{ma2024sit}.
In all experiments, we discretize $t\in[0,1]$ using a uniform grid $t_k=k/N$ with $N=1,000$ and integrate the reverse-time SDE in Eq.~\eqref{eq:sit} using the Euler--Maruyama sampler (same integrator and time grid as the official SiT implementation~\cite{ma2024sit}).
We set $\omega_t=\sigma_t=t$ and evaluate $(\alpha_t,\sigma_t,\dot{\alpha}_t,\dot{\sigma}_t,\omega_t)$ at each $t_k$ to compute the guided score via Eq.~\eqref{eq:s_from_v} and update $\mathbf{x}_{t_k}$.

Unlike score-based generative models (SGMs), which directly learn the score function, SiT learns the velocity fields that describe both the direction and rate of data evolution. This formulation provides more explicit control of the transition from noise to data and supports sampling via either the probability flow ODE or the reverse SDE. By leveraging continuous-time interpolants, SiT achieves more accurate modeling of the denoising trajectory, leading to improved reconstruction fidelity compared with the original DiT framework.

 \subsection{Classifier-Free Guidance (CFG) in CDPIR}
\label{subsec:cfg}
We adopt standard classifier-free guidance (CFG) \cite{ho2022classifierfreediffusionguidance} to enable controllable conditional generation without an external classifier. 
Let $\theta$ denote the learnable parameters of the diffusion network. At the training stage, instead of computing gradients from a pretrained classifier~\cite{dhariwal2021diffusion}, CFG jointly trains
a conditional velocity predictor $\mathbf{v}_{\theta}(\mathbf{x},t;c)$ with conditioning signal $c$, where in this work the signal is represented as a token derived from the class label, and an unconditional predictor $\mathbf{v}_{\theta}(\mathbf{x},t;\varnothing)$ is obtained by randomly replacing $c$ with a null token $\varnothing$, referred to as \emph{null conditioning}. This random-drop-label in training stage is the mechanism of cross-distribution prior.

At sampling time, we use the standard CFG difference form
\begin{equation}
\label{v_cal}
\mathbf{v}^{(\mu)}_{\theta}(\mathbf{x},t;c)
= \mathbf{v}_{\theta}(\mathbf{x},t;\varnothing)
  + \mu\big[\mathbf{v}_{\theta}(\mathbf{x},t;c)-\mathbf{v}_{\theta}(\mathbf{x},t;\varnothing)\big],
\end{equation}
where $\mu\!\ge\!0$ is the guidance scale and larger $\mu$ strengthens conditioning. 

\red{To describe our working hypothesis about the roles encouraged by null
conditioning and conditional guidance during training, we introduce the
following interpretive notation:}
\begin{align}
\underbrace{\mathbf{v}_{\mathrm{inv}}(\mathbf{x},t)}_{\text{domain-invariant}}
  &= \mathbf{v}_{\theta}(\mathbf{x},t;\varnothing), \\
\underbrace{\mathbf{v}_{\mathrm{spec}}(\mathbf{x},t; c)}_{\text{domain-specific}}
  &= \mathbf{v}_{\theta}(\mathbf{x},t; c) - \mathbf{v}_{\theta}(\mathbf{x},t;\varnothing).
\end{align}
\red{Here, the terms ``domain-invariant'' and ``domain-specific'' are used as
an interpretive description of our working hypothesis about the roles
encouraged by null conditioning and conditional guidance during training.
In this view, $\mathbf{v}_{\mathrm{inv}}$ is intended to
describe the component that tends to emphasize structures shared across
training domains, whereas $\mathbf{v}_{\mathrm{spec}}$ describes the
conditional residual that may reflect domain-specific appearance modulation.
Empirically, our CFG ablations and OOD generalization results are consistent
with this interpretation.}

Hence, Eq.~\eqref{v_cal} can be written compactly as
$\mathbf{v}^{(\mu)}_{\theta}(\mathbf{x},t;c)
= \mathbf{v}_{\mathrm{inv}}(\mathbf{x},t) + \mu\,\mathbf{v}_{\mathrm{spec}}(\mathbf{x},t;c)$.
This guided velocity is not directly applied to the sampler. Following the unified stochastic interpolant formulation, it is first mapped to a guided score:
\begin{equation}
\label{eq:vel2score}
\mathbf{s}^{(\mu)}_{\theta}(\mathbf{x}, t; c)
= \frac{1}{\sigma_t}\,
  \frac{\alpha_t\,\mathbf{v}^{(\mu)}_{\theta}(\mathbf{x}, t; c) - \dot{\alpha}_t\,\mathbf{x}}
       {\dot{\alpha}_t\,\sigma_t - \alpha_t\,\dot{\sigma}_t},
\end{equation}
where $\alpha_t$ and $\sigma_t$ are the interpolant coefficients, and $\dot{\alpha}_t$ and $\dot{\sigma}_t$ are the their respective time derivatives. 
Then, sampling integrates the reverse-time SDE
\begin{equation}
\label{eq:reverse_sde}
d\mathbf{x}
=
\mathbf{v}^{(\mu)}_{\theta}(\mathbf{x}, t; c)\,dt
+ \tfrac{1}{2}\,\omega_t\,\mathbf{s}^{(\mu)}_{\theta}(\mathbf{x}, t; c)\,dt
+ \sqrt{\omega_t}\,d\bar{\mathbf{w}},
\end{equation}
with diffusion schedule $\omega_t=\sigma_t$ and reverse-time Wiener increment $d\bar{\mathbf{w}}$.
We set the guidance scale to $\mu=1$ in all experiments, implying $\mathbf{v}^{(\mu)}_{\theta}(\mathbf{x},t;c)=\mathbf{v}_{\theta}(\mathbf{x},t;c)$ at inference.

\begin{figure}
    \centering
    \includegraphics[width=0.48\textwidth]{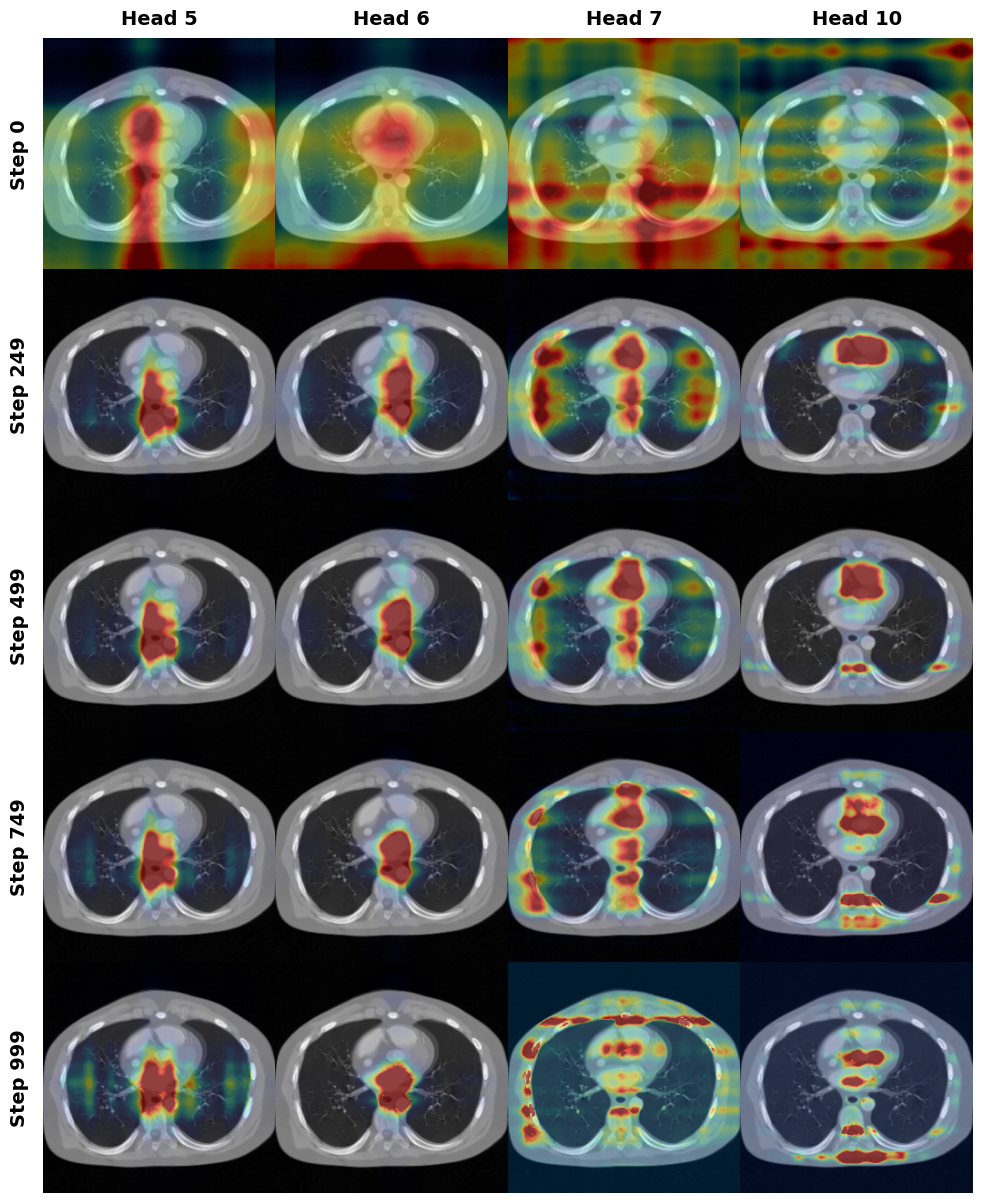}
    \caption{Representative attention-maps overlay from selected heads at different
sampling steps during reconstruction. The maps illustrate how the spatial
distribution of attention evolves across the sampling trajectory and differs
substantially across heads, indicating heterogeneous response patterns within
the transformer.}
    \label{fig:attention}
\end{figure}

\red{
To provide qualitative intuition
about  ``domain-invariant'' and ``domain-specific'', we visualize representative attention maps from several
heads across the sampling trajectory based on the sampling process of our model in Fig.~\ref{fig:attention}. We observe
that different heads may emphasize different aspects of the image during
sampling. For example, some heads (\emph{e.g.}, Head 7 in this example) exhibit
broader, anatomy-aligned response patterns, whereas others show more localized
or higher-frequency responses that may correlate with appearance-related features
such as texture, noise, and edge sharpness.}

\subsection{Model Architecture}

\begin{figure*}
  \centering
  \includegraphics[width=0.99\textwidth]{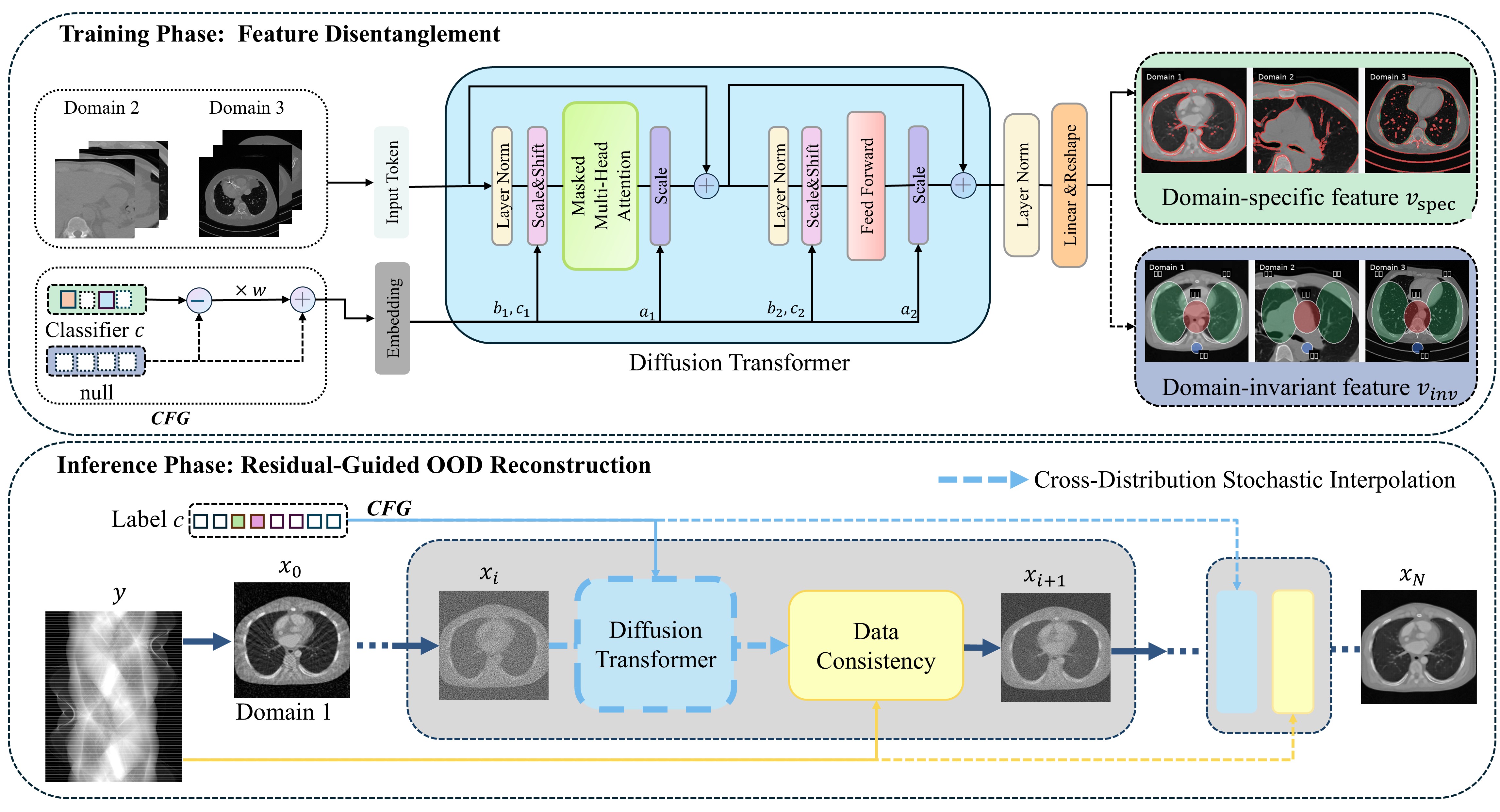}
  \caption{
\textbf{Architecture of the proposed domain-robust SVCT reconstruction framework.} 
The framework has two components. \textbf{Feature disentanglement during training}: A diffusion transformer, guided by classifier-free guidance, learns domain-invariant features that capture anatomical structures and domain-specific features that encode texture and noise to suppress sparse-view artifacts. \textbf{Residual-guided reconstruction during inference}: Cross-distribution diffusion priors are combined with iterative data consistency to refine reconstructions, yielding robust and artifact-free SVCT across heterogeneous domains.
}
  \label{fig:model}
\end{figure*}

Fig. \ref{fig:model} shows the overall architecture of our framework. The top part illustrates the training process based on CFG \cite{ho2022classifierfreediffusionguidance}, where the classifier $c$ organizes multiple datasets from different modalities, scanners, or pathologies. \red{By randomly dropping $c$ with $null$, the model learns both conditional and unconditional priors, which we interpret as encouraging two complementary representations.}
The first representation is the domain-invariant feature $v_{inv}$, obtained from the null branch. \red{In our interpretive view, this term tends to emphasize anatomy-level structures shared across training domains, including the heart, lungs, and spine, as well as consistent bone and soft-tissue regularities.} As shown in Fig. \ref{fig:model}, red highlights the cardiac region, green indicates the lungs, and blue represents the bones.  By modeling this anatomy-aware prior, $v_{inv}$ helps mitigate cross-domain distribution shifts and ensures robust generalization to unseen domains.
The second representation is the domain-specific feature $v_{spec}$, derived from the conditional branch. \red{In our interpretive view, this term reflects domain-specific appearance characteristics.} As illustrated in Fig.~\ref{fig:model}, these responses appear as edge-like patterns that may be associated with fine-scale information such as boundary sharpness, noise
characteristics, and texture details. By modeling these priors, $v_{spec}$ helps suppress sparse-view artifacts while preserving domain-consistent high-frequency details crucial for realistic reconstruction.

The bottom part in Fig.~\ref{fig:model} shows the inference phase for residual-guided OOD SVCT reconstruction. Starting from the initialization image $\mathbf{x}_0$ obtained via reconstruction of measurement $\mathbf{y}$, the CT image is iteratively reconstructed through intermediate steps (\( \mathbf{x}_i \)) until the final reconstruction (\( \mathbf{x}_N \)) is achieved.
Each iteration combines a diffusion transformer denoiser and a data consistency module. The denoiser leverages cross-distribution priors with Cross-Distribution Stochastic Interpolation, providing a flexible interpolant that stabilizes sampling and improves robustness to domain shifts, while the data consistency module enforces fidelity to measured projections for accurate reconstructions. In this work, ASD-POCS is employed as the data consistency method and initialization method, enhancing the accuracy of the reconstruction process.

\subsection{Residual-Guided Alternating Solver}
This approach defines the minimization problem as follows  
\begin{equation}
\min_{\mathbf{x}} J(\mathbf{x}) + \lambda_1 \|\nabla \mathbf{x}\|_1, 
\quad \text{s.t.} \quad \|\mathbf{A} \mathbf{x} - \mathbf{y}\|_2^2 \leq \delta_n,
\label{ObjFunc1}
\end{equation}
where $J(\mathbf{x})$ denotes the objective of the reverse diffusion process, $\|\nabla \mathbf{x}\|_1$ is the Total Variation (TV) regularization with $\lambda_1$ being the weight, and $\delta_n$ specifies the noise tolerance. This constraint enforces consistency with the measurement data while guiding the diffusion prior.  

To solve Eq.~(\ref{ObjFunc1}), it can be reformulated into an unconstrained problem:  
\begin{equation}
\min_{\mathbf{x}} \|A \mathbf{x} - \mathbf{y}\|_2^2 
+ \lambda_1 \|\nabla \mathbf{x}\|_1 
+ \lambda_2 J(\mathbf{x}),
\label{ObjFunc2}
\end{equation}
where $\lambda_1$ and $\lambda_2$ balance the data fidelity, TV, and diffusion prior. Introducing an auxiliary variable $\mathbf{r}$ yields an equivalent constrained problem:  
\begin{equation}
\min_{\mathbf{x}, \mathbf{r}} 
\|A \mathbf{x} - \mathbf{y}\|_2^2 
+ \lambda_1 \|\nabla \mathbf{x}\|_1 
+ \lambda_2 J(\mathbf{r}), 
\quad \text{s.t.} \quad \mathbf{r} = \mathbf{x}.
\label{ObjFunc3}
\end{equation}
Eq.(\ref{ObjFunc3}) can be solved by alternatively solving two sub-problems: 
\begin{equation}
\mathbf{r}_{i+1} = \arg\min_{\mathbf{r}} 
\|\mathbf{r} - \mathbf{x}_i\|_2^2 + \lambda_2 J(\mathbf{r}), 
\label{GenerativeSubproblem}
\end{equation}
\begin{equation}
\mathbf{x}_{i+1} = \arg\min_{\mathbf{x}} 
\|A \mathbf{x} - \mathbf{y}\|_2^2 
+ \lambda_1 \|\nabla \mathbf{x}\|_1 
+ \lambda_3 \|\mathbf{x} - \mathbf{r}_{i+1}\|_2^2, 
\label{ConsistencySubproblem}
\end{equation}
where $i$ denotes the iteration step, and $\lambda_3$ is to control $\|\mathbf{x} - \mathbf{r}_{i+1}\|_2^2$.

Diffusion Generative Subproblem Eq.(\ref{GenerativeSubproblem}) optimizes the diffusion model's objective function while trying to minimize the difference between $\mathbf{r}$ and $\mathbf{x}_i$. In our model, this problem is solved by Stochastic Euler-Maruyama Sampler based on Eq.(\ref{eq:reverse_sde}). Specifically, it will compute the drift and diffusion components using the classifier-free guidance score $\mathbf{s}^{(\mu)}_{\theta}$ and velocity $\mathbf{v}^{(\mu)}_{\theta}$.

To solve the data-fidelity subproblem Eq.(\ref{ConsistencySubproblem}), we adopt the ASD-POCS framework with iTV minimization~\cite{ritschl2011improved}.
We use $i$ to denote the outer CDPIR iteration, and $p$ to denote the inner ASD-POCS iteration to solve the data-fidelity subproblem within each outer iteration.
Given the current estimate $\mathbf{x}^{i}$, we first run the diffusion prior step to obtain $\mathbf{r}^{i+1}$.
Then, we solve the data-fidelity subproblem by initializing ASD-POCS with $\mathbf{r}^{i+1}$ and performing $P$ iTV-regularized ASD-POCS iterations.
In inner iteration $p$, ASD-POCS computes a projection update using SART (Simultaneous Algebraic Reconstruction Technique), and we denote the resulting intermediate image as $X_{\mathrm{SART}}^{i,p+1}$. To accelerate the projection update, we employ Ordered-Subset SART (OS-SART) and use $K$ to denote the number of subsets.

Next, $Q$ iTV diffusion steps are applied to reduce the TV norm and obtain the regularized image $X_{\mathrm{TV}}^{i,p+1,Q}$.
The final inner-loop update is computed by an adaptive fusion
\begin{equation}
\mathbf{x}^{i,p+1}
=
X_{\mathrm{SART}}^{i,p+1}
+
\lambda_{\mathrm{TV}}
\Bigl[
X_{\mathrm{TV}}^{i,p+1,Q}
-
X_{\mathrm{SART}}^{i,p+1}
\Bigr],
\label{eq:lambda_combine}
\end{equation}
where $\lambda_{\mathrm{TV}}\in [0,1]$ is chosen adaptively to control projection-domain consistency.
Following~\cite{ritschl2011improved}, $\lambda_{\mathrm{TV}}$ is determined by enforcing
\begin{equation}
\bigl\lVert A\mathbf{x}^{i,p+1}-\mathbf{y}\bigr\rVert_2^2
=
\epsilon_{\mathrm{SART}}^{i,p+1}
+
w\Bigl[
\epsilon^{i,p}
-
\epsilon_{\mathrm{SART}}^{i,p+1}
\Bigr],
\label{eq:lambda_condition}
\end{equation}
where $w=0.8$ is fixed, $\epsilon_{\mathrm{SART}}^{i,p+1}=\lVert AX_{\mathrm{SART}}^{i,p+1}-\mathbf{y}\rVert_2^2$ is the residual after the SART update, and $\epsilon^{i,p}=\lVert A\mathbf{x}^{i,p}-\mathbf{y}\rVert_2^2$ is the residual from the previous inner iteration.

As implemented in our ASD-POCS solver, Eq.(\ref{eq:lambda_condition}) yields a quadratic equation in $\lambda_{\mathrm{TV}}$, and we select the valid root and clip it to $[0,1]$ for numerical stability.
The complete procedure is summarized in Algorithm~\ref{alg:cdpir}.

\begin{algorithm}
\caption{CDPIR: alternating diffusion prior and ASD-POCS iTV data consistency}
\label{alg:cdpir}
\begin{algorithmic}[1]
\Require Measurements $\mathbf{y}$, system matrix $A$, initial estimate $\mathbf{x}^{0}$, Outer iterations $I$, reverse SDE steps $N$, ASD-POCS iterations $P$, OS-SART subsets $K$, iTV steps $Q$, Time grid $\{t_n\}_{n=0}^{N}$, schedules $\alpha_{t_n}$, $\sigma_{t_n}$, $\dot{\alpha}_{t_n}$, $\dot{\sigma}_{t_n}$, diffusion coefficient $\omega_{t_n}$
\For{$i=0$ to $I-1$}
    \State \textbf{Diffusion prior step}
    \State $\mathbf{z}_{N} \gets \mathbf{x}^{i}$
    \For{$n=N$ down to $1$}
        \State $h \gets t_{n}-t_{n-1}$
        \State $\mathbf{v}_n \gets \mathbf{v}_\theta\{\mathbf{z}_n,t_n;c\}$ \Comment{Velocity field, Eq.~\eqref{v_cal}}
        \State $\mathbf{s}_n \gets \textsc{Score}\{\mathbf{v}_n,\mathbf{z}_n,\alpha_{t_n},\sigma_{t_n},\dot{\alpha}_{t_n},\dot{\sigma}_{t_n}\}$ \Comment{Eq.~\eqref{eq:vel2score}}
        \State Sample $\boldsymbol{\epsilon}_n \sim \mathcal{N}\{\mathbf{0},\mathbf{I}\}$
        \State $\mathbf{z}_{n-1} \gets \mathbf{z}_n + h\Bigl[\mathbf{v}_n + \tfrac{1}{2}\omega_{t_n}\mathbf{s}_n\Bigr] + \sqrt{\omega_{t_n}h}\,\boldsymbol{\epsilon}_n$ \Comment{Eq.~\eqref{eq:reverse_sde}}
    \EndFor
    \State $\mathbf{r}^{i+1} \gets \mathbf{z}_0$
    \State \textbf{Data-fidelity step via ASD-POCS iTV}
    \State $\mathbf{x}^{i,0} \gets \mathbf{r}^{i+1}$
    \For{$p=0$ to $P-1$}
        \State $X_{\mathrm{SART}}^{i,p+1} \gets \mathrm{OSSART}\{\mathbf{x}^{i,p},\mathbf{y},A,K\}$ \Comment{Projection update}
        \State $X_{\mathrm{TV}}^{i,p+1,Q} \gets \mathrm{iTV}\{X_{\mathrm{SART}}^{i,p+1},Q\}$ \Comment{TV regularization~\cite{ritschl2011improved}}
        \State Solve Eq.~\eqref{eq:lambda_condition} for $\lambda_{\mathrm{TV}}$, and Clip $\lambda_{\mathrm{TV}}$ to $[0,1]$
        \State $\mathbf{x}^{i,p+1} \gets X_{\mathrm{SART}}^{i,p+1} + \lambda_{\mathrm{TV}}\Bigl[X_{\mathrm{TV}}^{i,p+1,Q}-X_{\mathrm{SART}}^{i,p+1}\Bigr]$ \Comment{Eq.~\eqref{eq:lambda_combine}}
    \EndFor
    \State $\mathbf{x}^{i+1} \gets \mathbf{x}^{i,P}$
\EndFor
\State \Return $\mathbf{x}^{I}$
\end{algorithmic}
\end{algorithm}

\section{Experiment and Result}

\subsection{Experimental Details and Parameters}

For CDPIR, we adopt the SiT backbone in the Big configuration (Table~\ref{tab:model_size}) and remove the autoencoder used in the original SiT/DiT pipeline so that both training and inference are performed entirely in the \emph{image domain}. This choice avoids latent-space compression artifacts and enables a more direct integration with projection-domain data-consistency updates for SVCT.

\begin{table}[htbp]
    \centering
    \caption{Model capacity comparison for CDPIR (Patch size = 2).}
    \resizebox{\columnwidth}{!}{
        \begin{tabular}{lcccc}
            \toprule
            Model & Depth & Hidden Size & Attention Heads & Parameters (M) \\
            \midrule
            CDPIR-B-2 (Big)   & 12 & 768 & 12 & 142.9 \\
            CDPIR-S-2 (Small) & 12 & 384 & 6  & 39.1  \\
            \bottomrule
        \end{tabular}
    }
    \label{tab:model_size}
\end{table}
Unless otherwise stated, CDPIR uses a non-overlapping patch tokenization with patch size $p{=}2$ and the SiT-B architecture (depth $L{=}12$, hidden size $D{=}768$, and $H{=}12$ attention heads), resulting in $\sim$142.8M parameters. We implement CFG using domain tokens: $c{=}0$ for AAPM and $c{=}1$ for COCA, while the conditioning token is randomly dropped to a dedicated null token with a fixed probability $p_{\mathrm{drop}}$ during training. At inference, we use the standard CFG with guidance scale $\mu{=}1.0$ unless otherwise specified. Training hyperparameters follow DiT~\cite{peebles2023scalable}: AdamW with learning rate $1{\times}10^{-4}$, $\beta{=}(0.9,0.999)$ (weight decay $0$), together with an EMA model for evaluation. During inference, we use $N{=}1,000$ diffusion steps. For physics-based refinement, we employ ASD-POCS as the data-consistency module with fixed parameters (outer iterations $P{=}10$, TV step $Q{=}1$, and OS-SART subsets $K{=}5$), and we use 30 iterations for initialization.

For OOD sparse-view reconstruction tasks, we benchmark our method against several the state-of-the-art approaches, including MCG~\cite{chung2022improving}, DiffusionMBIR~\cite{chung2023solving}, and DDS~\cite{chung2023decomposed}. We also perform qualitative comparisons with ASD-POCS to demonstrate how learned priors from diffusion models improve reconstruction performance. To ensure fair and consistent comparisons, we strictly follow the original training configurations and pretrained models released by the authors of all baseline methods. We do not modify network architectures, parameter sizes, loss functions, or training schedules of these methods. All approaches are evaluated using the same forward models and testing datasets, with inference-time settings adjusted only when required by the original implementations. All experiments were conducted on a single NVIDIA RTX~A6000 (48GB) GPU.

\subsection{Datasets and Evaluation}

We conduct SVCT reconstruction experiments using five datasets:

\begin{enumerate}
    \item \textbf{AAPM 2016 Low-Dose CT Grand Challenge Dataset}: This publicly available dataset provides clinical CT scans acquired on Siemens scanners under low-dose protocols. It contains 5,410 slices in total, corresponding to 9 patients. Following our split, 4,850 slices from 8 patients are used for training, and 560 slices from 1 patient are reserved for testing.

    \item \textbf{Stanford AIMI COCA Dataset}~\cite{aimi2022coca}: To evaluate reconstruction performance under diverse cardiac imaging conditions, we use the COCA dataset curated by the Stanford Center for Artificial Intelligence in Medicine and Imaging. It includes a wide range of cardiac CT scans acquired with varying protocols. In our study, we use 4,210 slices from 83 patients for training and 418 slices from 11 patients for testing.

    \item \textbf{XCAT Simulated Cardiac Dataset}~\cite{segars20104d}: This dataset is generated with the 4D Extended Cardiac-Torso (XCAT) phantom~\cite{segars20104d}, developed at Duke University. Projection data simulate a GE CT scanner with a source-to-detector distance of 950~mm, a source-to-object distance of 538.5~mm, and a curved detector array of 835 elements across 984 views. The field of view diameter is 500~mm, and each cardiac cycle is divided into 200 time phases per second. Clinical noise is approximated with a tube current-time product of 0.25~mAs per projection. The dataset comprises 10 simulated patients, each contributing 352 reconstructed slices at $256 \times 256$ resolution, totaling 3,520 images. A leave-one-out cross-validation strategy is used, training on 9 patients and testing on the remaining 1 patient for each fold.

    \item \textbf{GE Clinical Cardiac Dataset}: This dataset consists of high-resolution cardiac CT scans acquired from patients using a 256-slice GE scanner under IRB approval at Vanderbilt University Medical Center and the University of Massachusetts Lowell. In this study, we evaluate on 1 clinical test case. Scans were acquired in axial mode with clockwise gantry rotation. The source-to-isocenter and source-to-detector distances were 625.61~mm and 1,097.6~mm, respectively. A curved detector array with 828 row-wise cells of $1.09 \times 1.09$~mm$^2$ was used to capture 984 projections over 360$^\circ$. Reconstructions were performed with the Feldkamp--Davis--Kress (FDK) algorithm under cone-beam geometry.

    \item \textbf{MARS PCCT Extremity Dataset}: Patients aged 21 years and older referred from the fracture clinic were recruited for this clinical trial. In this study, we evaluate on 1 clinical test case. Wrist images were acquired using the MARS Extremity photon-counting CT (PCCT) scanner at 120~kVp in helical mode, with a tube current of 61~$\mu$A and exposure time of 160~ms. The Medipix3RX detector consists of 12 chips arranged into three flat panels approximating an arc. Each projection has 128 rows and 1,547 columns with a pixel size of $0.11\times0.11$~mm$^2$. The source-to-isocenter and source-to-detector distances were 199.52~mm and 271.88~mm, respectively. Data were collected in 8 energy bins; in this work, we use the $8^{\mathrm{th}}$ bin. The 3D helical cone-beam projections are first rebinned into 2D fan-beam sinograms, followed by equal-distance FBP reconstruction to generate ground-truth CT images.
\end{enumerate}

For quantitative evaluation, we adopt peak signal-to-noise ratio (PSNR), structural similarity index measure (SSIM), and the Learned Perceptual Image Patch Similarity (LPIPS) metric, where higher PSNR/SSIM and lower LPIPS indicate better image quality. 
\red{
In the clinical experiments, including GE and MARS PCCT, all quantitative metrics are computed within a fixed circular FOV defined for each case to exclude background regions. Within each case, the same mask is applied to all compared methods to ensure a fair comparison. All visualizations use a fixed HU window per dataset that is applied uniformly to all methods.
}

\subsection{Numerical Simulation Results}
\subsubsection{Dataset Preparation} 
In  numerical simulation experiments, projection data for AAPM, COCA, and XCAT datasets are generated using CT system parameters derived from the GE clinical cardiac dataset described above using LEAP toolbox\cite{kim2023differentiableforwardprojectorxray}. 
Sparse-view inputs are generated by uniformly downsampling the full-view projections. Specifically, for all three datasets, 55 views are uniformly sampled from the original 984 views to simulate sparse-view acquisition conditions.

\subsubsection{OOD Qualitative Analysis}
To evaluate robustness under domain shift, we train all methods on the AAPM dataset and evaluate them on the XCAT phantom dataset, which differs from the training distribution in both anatomical appearance and noise/texture statistics.  For clinical interpretability, chest CT results are displayed in both the lung window ([-1000, 150]HU) and the mediastinal window ([-160, 240]HU), providing complementary visualization of parenchymal and soft-tissue structures. Consistent trends are observed in both windows, where CDPIR maintains clearer anatomical delineation with fewer artifacts.
As shown in Fig.~3, CDPIR yields reconstructions that are visually closer to the ground truth than competing baselines.
\begin{figure*}
    \centering
    \includegraphics[width=0.98\textwidth]{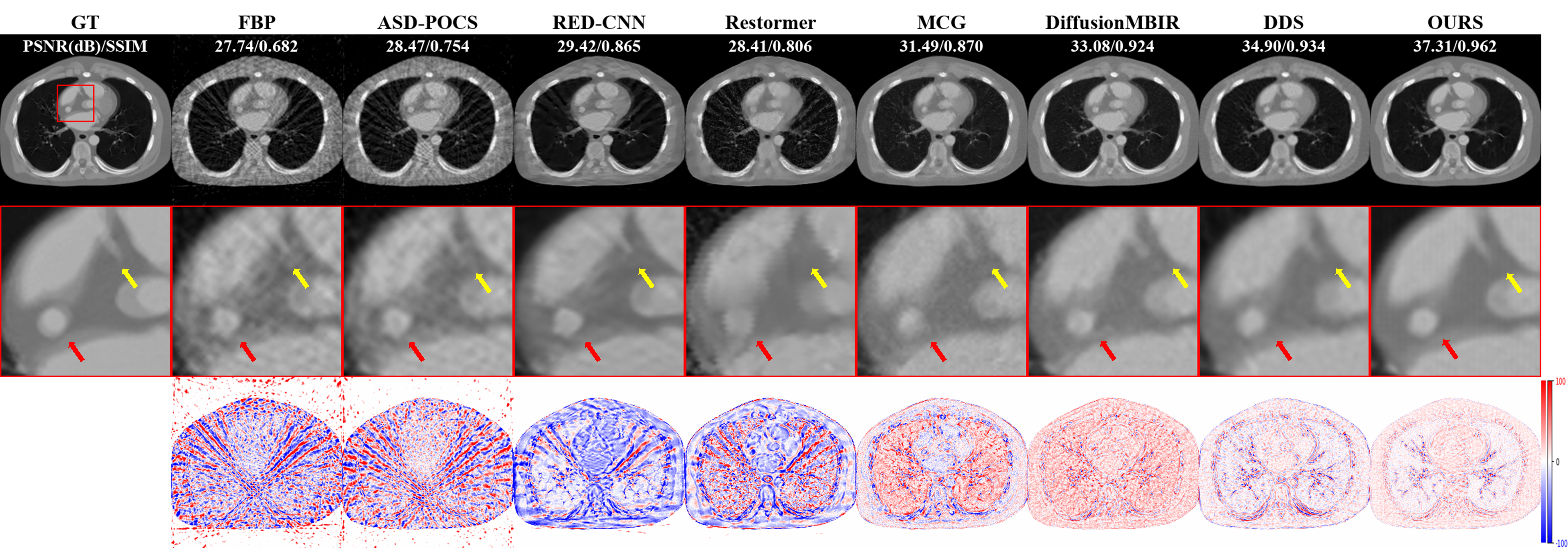}
    \caption{\textbf{OOD visualization of a representative image slice in XCAT dataset.} The first row displays the full reconstruction (HU), followed by the error maps (second row) which highlight the residual noise relative to the Ground Truth (GT). The third and fourth rows present the Lung window ([-1000, 150] HU) and its corresponding magnified ROI, while the fifth and sixth rows show the Mediastinal window ([-160,240] HU) and its ROI. The yellow arrows highlight high-contrast edge preservation, while red arrows indicate superior recovery of low-contrast textures. All full-slice images are displayed using a window setting of $[-1000, 750]$ HU unless otherwise specified.}
    \label{fig:xcat}
\end{figure*}

\subsubsection{In-Distribution Qualitative Analysis}
We evaluate in-distribution performance by training and testing all methods on the COCA cardiac CT dataset. As shown in Fig.~\ref{fig:coca}, CDPIR produces reconstructions that more closely match the ground truths (GT) in both global appearance and fine anatomical details compared with competing baselines.
The ROI magnifications emphasize coronary calcifications (red arrows), which require accurate preservation of high-contrast boundaries while avoiding artificial sharpening. CDPIR maintains clearer calcification contours and surrounding soft-tissue textures, whereas several baselines exhibit either residual streaking/noise or over-smoothing of local structures. The error maps further confirm these observations: CDPIR shows reduced residual artifacts and fewer structured errors around the highlighted regions, indicating improved fidelity under the in-distribution setting.

\begin{figure*}
    \centering
    \includegraphics[width=0.95\textwidth]{figure2.jpg}
    \caption{\textbf{In-distribution visualization of a representative image slice in COCA dataset. }
    The number of views and layout are the same as in Fig.~\ref{fig:xcat}. 
    The display  window is [-1000, 1000] HU}
    \label{fig:coca}
\end{figure*}
\newcommand{\gvar}[1]{{\color{gray}\scriptsize\raisebox{1pt}{$\pm$#1}}}

\begin{table*}[t]
\centering
\caption{Quantitative Comparison of Models Trained on \textbf{AAPM}. Best results are in \textbf{bold}; variances in \textcolor{gray}{gray}. }
\label{tab:train_aapm}
\setlength{\tabcolsep}{3pt}
\renewcommand{\arraystretch}{1.2}
\scriptsize
\begin{tabular}{l | ccc | ccc | ccc}
\toprule
\textbf{Method} &
\multicolumn{3}{c|}{\textbf{AAPM (ID)}} &
\multicolumn{3}{c|}{\textbf{COCA (OOD)}} &
\multicolumn{3}{c}{\textbf{XCAT (OOD)}} \\
\cmidrule(lr){2-4} \cmidrule(lr){5-7} \cmidrule(lr){8-10}
 &
\textbf{PSNR}$\uparrow$ & \textbf{SSIM}$\uparrow$ & \textbf{LPIPS}$\downarrow$ &
\textbf{PSNR}$\uparrow$ & \textbf{SSIM}$\uparrow$ & \textbf{LPIPS}$\downarrow$ &
\textbf{PSNR}$\uparrow$ & \textbf{SSIM}$\uparrow$ & \textbf{LPIPS}$\downarrow$ \\
\midrule

ASD-POCS
& 26.27 \gvar{3.18} & 0.741 \gvar{0.024} & 0.198 \gvar{0.020}
& 28.32 \gvar{3.19} & 0.820 \gvar{0.049} & 0.182 \gvar{0.027}
& 31.62 \gvar{1.21} & 0.826 \gvar{0.013} & 0.175 \gvar{0.010} \\
\midrule

RED-CNN
& 35.48 \gvar{1.17} & 0.902 \gvar{0.000} & 0.062 \gvar{0.000}
& 28.43 \gvar{2.51} & 0.764 \gvar{0.002} & 0.210 \gvar{0.004}
& 32.50 \gvar{0.46} & 0.876 \gvar{0.002} & 0.075 \gvar{0.000} \\

Restormer
& 34.38 \gvar{1.41} & 0.889 \gvar{0.001} & 0.066 \gvar{0.000}
& 28.35 \gvar{1.79} & 0.770 \gvar{0.001} & 0.222 \gvar{0.002}
& 30.19 \gvar{0.18} & 0.809 \gvar{0.000} & 0.132 \gvar{0.000} \\

MCG
& 32.71 \gvar{0.99} & 0.842 \gvar{0.000} & 0.029 \gvar{0.004}
& 30.72 \gvar{2.49} & 0.804 \gvar{0.034} & 0.171 \gvar{0.034}
& 29.86 \gvar{0.91} & 0.701 \gvar{0.000} & 0.083 \gvar{0.010} \\

DiffusionMBIR
& 33.79 \gvar{1.97} & 0.901 \gvar{0.008} & 0.016 \gvar{0.002}
& 29.91 \gvar{0.19} & 0.842 \gvar{0.016} & 0.152 \gvar{0.028}
& 31.46 \gvar{1.86} & 0.767 \gvar{0.092} & 0.082 \gvar{0.073} \\

DDS
& 32.22 \gvar{1.98} & 0.881 \gvar{0.001} & 0.028 \gvar{0.080}
& 36.01 \gvar{2.62} & 0.937 \gvar{0.144} & 0.068 \gvar{0.028}
& 33.60 \gvar{1.09} & 0.887 \gvar{0.054} & 0.087 \gvar{0.006} \\

DDIP3D
& 34.78 \gvar{4.46} & 0.908 \gvar{0.000} & 0.021 \gvar{0.000}
& 36.71 \gvar{6.32} & 0.939 \gvar{0.000} & \textbf{0.030} \gvar{0.000}
& 36.32 \gvar{6.45} & 0.885 \gvar{0.000} & \textbf{0.012} \gvar{0.000} \\
\midrule

\textbf{CDPIR (Ours)}
& \textbf{40.19} \gvar{1.59} & \textbf{0.963} \gvar{0.008} & \textbf{0.019} \gvar{0.006}
& \textbf{40.42} \gvar{3.23} & \textbf{0.963} \gvar{0.026} & 0.037 \gvar{0.003}
& \textbf{38.36} \gvar{0.75} & \textbf{0.953} \gvar{0.003} & 0.018 \gvar{0.003} \\
\bottomrule
\end{tabular}
\end{table*}

\begin{table*}[t]
\centering
\caption{Quantitative Comparison of Models Trained on \textbf{COCA}. Best results are in \textbf{bold}; variances in \textcolor{gray}{gray}.}
\label{tab:train_coca}
\setlength{\tabcolsep}{3pt}
\renewcommand{\arraystretch}{1.2}
\scriptsize
\begin{tabular}{l | ccc | ccc | ccc}
\toprule
\textbf{Method} &
\multicolumn{3}{c|}{\textbf{AAPM (OOD)}} &
\multicolumn{3}{c|}{\textbf{COCA (ID)}} &
\multicolumn{3}{c}{\textbf{XCAT (OOD)}} \\
\cmidrule(lr){2-4} \cmidrule(lr){5-7} \cmidrule(lr){8-10}
 &
\textbf{PSNR}$\uparrow$ & \textbf{SSIM}$\uparrow$ & \textbf{LPIPS}$\downarrow$ &
\textbf{PSNR}$\uparrow$ & \textbf{SSIM}$\uparrow$ & \textbf{LPIPS}$\downarrow$ &
\textbf{PSNR}$\uparrow$ & \textbf{SSIM}$\uparrow$ & \textbf{LPIPS}$\downarrow$ \\
\midrule

ASD-POCS
& 26.27 \gvar{3.18} & 0.741 \gvar{0.024} & 0.198 \gvar{0.020}
& 28.32 \gvar{3.19} & 0.820 \gvar{0.049} & 0.182 \gvar{0.027}
& 31.62 \gvar{1.21} & 0.826 \gvar{0.013} & 0.175 \gvar{0.010} \\
\midrule

RED-CNN
& 29.05 \gvar{0.43} & 0.794 \gvar{0.000} & 0.148 \gvar{0.000}
& 33.03 \gvar{2.20} & 0.876 \gvar{0.001} & 0.124 \gvar{0.003}
& 33.34 \gvar{0.30} & 0.876 \gvar{0.000} & 0.068 \gvar{0.000} \\

Restormer
& 28.71 \gvar{0.34} & 0.758 \gvar{0.000} & 0.174 \gvar{0.001}
& 32.79 \gvar{1.99} & 0.856 \gvar{0.000} & 0.100 \gvar{0.002}
& 31.07 \gvar{0.19} & 0.806 \gvar{0.000} & 0.131 \gvar{0.000} \\

MCG
& 28.54 \gvar{4.15} & 0.730 \gvar{0.001} & 0.096 \gvar{0.018}
& 32.77 \gvar{3.45} & 0.852 \gvar{0.001} & 0.079 \gvar{0.026}
& 29.86 \gvar{0.91} & 0.782 \gvar{0.000} & 0.085 \gvar{0.008} \\

DiffusionMBIR
& 33.58 \gvar{2.66} & 0.880 \gvar{0.000} & 0.041 \gvar{0.002}
& 34.22 \gvar{2.49} & 0.911 \gvar{0.022} & 0.048 \gvar{0.016}
& 31.84 \gvar{1.36} & 0.780 \gvar{0.040} & 0.054 \gvar{0.009} \\

DDS
& 32.86 \gvar{4.24} & 0.861 \gvar{0.001} & 0.030 \gvar{0.037}
& 34.16 \gvar{2.55} & 0.924 \gvar{0.004} & 0.075 \gvar{0.014}
& 34.55 \gvar{2.46} & 0.925 \gvar{0.004} & 0.069 \gvar{0.006} \\

DDIP3D
& 33.53 \gvar{3.72} & 0.905 \gvar{0.000} & 0.041 \gvar{0.000}
& 35.04 \gvar{9.12} & 0.926 \gvar{0.000} & 0.043 \gvar{0.000}
& 33.82 \gvar{2.24} & 0.829 \gvar{0.028} & 0.021 \gvar{0.001} \\
\midrule

\textbf{CDPIR (Ours)}
& \textbf{39.82} \gvar{1.76} & \textbf{0.963} \gvar{0.008} & \textbf{0.020} \gvar{0.006}
& \textbf{39.74} \gvar{3.01} & \textbf{0.951} \gvar{0.001} & \textbf{0.030} \gvar{0.009}
& \textbf{40.06} \gvar{1.16} & \textbf{0.971} \gvar{0.000} & \textbf{0.012} \gvar{0.002} \\
\bottomrule
\end{tabular}
\end{table*}

\begin{figure*}[!t]
    \centering
    \includegraphics[width=\linewidth]{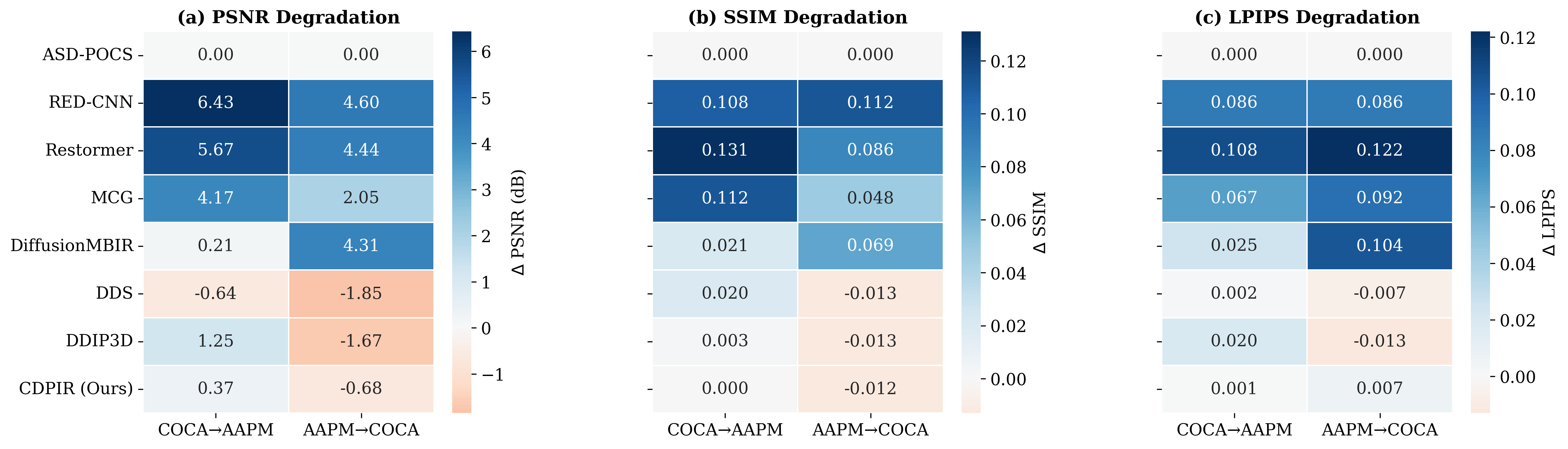}
    \caption{\textbf{Fixed-test OOD degradation heatmaps for PSNR, SSIM, and LPIPS across different transfer scenarios.} The degradation metric ($\Delta_{\text{OOD}}$) is formulated such that greater positive values consistently indicate more severe performance degradation under domain shift. Conversely, values near zero indicate robust generalization, and negative values denote performance gains compared to the ID baseline.}
    \label{fig:ood_heatmaps}
\end{figure*}

\subsubsection{Quantitative Performance Across Datasets}

We evaluate cross-domain generalizability by training CDPIR and all baselines on two source domains AAPM and COCA and testing on three target domains AAPM, COCA, and XCAT, covering both ID and OOD settings. Tables~\ref{tab:train_aapm} and~\ref{tab:train_coca} report PSNR, SSIM, and LPIPS for each train--test configuration. Testing on the training domain is treated as ID, while the remaining targets are treated as OOD.

Because absolute metrics across heterogeneous datasets are strongly affected by intrinsic domain difficulty, such as acquisition dose and noise level, anatomy complexity, and field-of-view, raw PSNR and SSIM are not directly comparable across datasets and can occasionally yield higher absolute scores in OOD cases. To quantify OOD degradation without this factor, we additionally measure the drop under a \textbf{fixed testing domain}. For a given testing dataset $D$, we treat ID as Training $D\rightarrow$Testing $D$ and OOD as Training $D'\neq D\rightarrow$Testing $D$. We define degradation such that $\Delta>0$ consistently indicates worse performance:
\begin{align}
\Delta^{\mathcal{M}}_{\text{OOD}}(D'\!\rightarrow\!D)=
\begin{cases}
\mathcal{M}_{D\rightarrow D}-\mathcal{M}_{D'\rightarrow D}, & \text{PSNR/SSIM},\\[2pt]
\mathcal{M}_{D'\rightarrow D}-\mathcal{M}_{D\rightarrow D}, & \text{LPIPS},
\end{cases}
\label{eq:fixed_test_drop}
\end{align}
where $\mathcal{M}_{A\rightarrow B}$ denotes the dataset-average metric on the testing dataset $B$ produced by a model trained on dataset $A$. LPIPS is inverted because lower values indicate better perceptual similarity. We visualize $\Delta^{\mathcal{M}}_{\text{OOD}}$ as heatmaps in Fig.~\ref{fig:ood_heatmaps} for the two fixed-test transfer scenarios COCA$\rightarrow$AAPM and AAPM$\rightarrow$COCA. Values near zero indicate preserved performance under training-domain shift.
Take an example of AAPM$\rightarrow$COCA, the fixed testing domain is $D=\text{COCA}$, so the ID reference is $\mathcal{M}_{\text{COCA}\rightarrow\text{COCA}}$, while the OOD-trained result is $\mathcal{M}_{\text{AAPM}\rightarrow\text{COCA}}$.
For instance, for RED-CNN, $\Delta^{\text{PSNR}}_{\text{OOD}}(\text{AAPM}\!\rightarrow\!\text{COCA})=
33.03-28.43=4.60$ dB and 
$\Delta^{\text{LPIPS}}_{\text{OOD}}(\text{AAPM}\!\rightarrow\!\text{COCA})=
0.210-0.124=0.086$, indicating a clear degradation when training shifts from COCA to AAPM under the same COCA testing set.
In contrast, for CDPIR, $\Delta^{\text{PSNR}}_{\text{OOD}}(\text{AAPM}\!\rightarrow\!\text{COCA})=
39.74-40.42=-0.68$ dB (\emph{i.e.}, no drop in PSNR).

Across both training settings, CDPIR achieves strong absolute reconstruction quality and exhibits markedly improved robustness under the fixed-test evaluation. In Fig.~\ref{fig:ood_heatmaps}, supervised CNN and Transformer baselines show substantial degradation under training-domain shift, with multi-dB drops in $\Delta$PSNR, noticeable decreases in $\Delta$SSIM, and consistent perceptual degradation in LPIPS, indicating that their improvements are tightly coupled to the training distribution. In contrast, CDPIR shows near-zero degradation across transfer scenarios, demonstrating substantially more stable generalization when the training domain changes while keeping the same testing domain.

\subsection{Reconstruction on GE Clinical Cardiac Dataset }
\begin{figure*}
    \centering
    \includegraphics[width=1\textwidth]{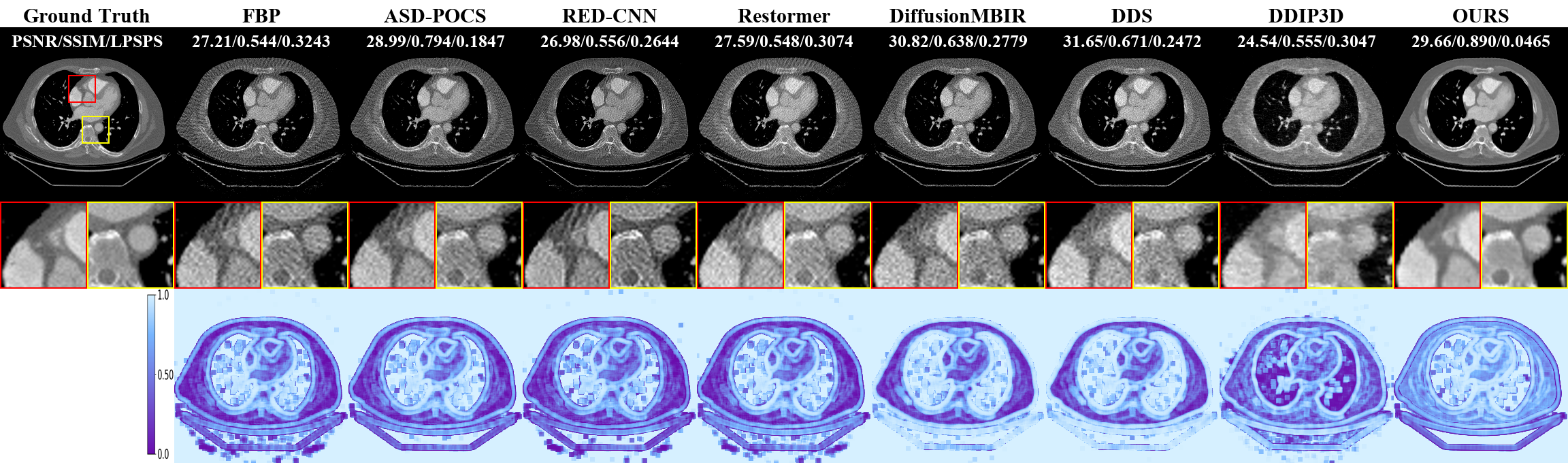}
    \caption{\textbf{Zero-shot reconstruction on the Clinical GE dataset.} Models are trained on AAPM and tested on 123-view clinical cardiac data. Iterative methods outperform U-Net diffusion models, but all baselines show severe sparse-view artifacts. Our method suppresses artifacts and achieves the highest SSIM (+0.07 over DDS), demonstrating strong generalization under extreme sparsity, HU[-800,320]. Note that the bottom row is the SSIM map.All methods are visualized with identical windowing and normalization for fair comparison.}
    \label{fig:ge}
\end{figure*}

For the GE clinical cardiac experiments, raw projections are available and fan-beam sinograms are extracted. Ground-truth CT images are obtained by reconstructing from the full-view (984-view) sinograms. We evaluate in a \emph{zero-shot} setting without any pretraining or fine-tuning on clinical data to better reflect real-world deployment: all diffusion models and CDPIR are trained on AAPM only, and the sparse-view clinical sinograms are used as the sole input at test time. To simulate extreme sparsity, we uniformly subsample 123 views from the original 984-view acquisition. For evaluation, we reconstruct the full 3D volume slice-by-slice; Fig.~\ref{fig:ge} shows the central slice as a representative example.

As shown in Fig.~\ref{fig:ge}, classical iterative reconstruction (e.g., ASD-POCS) outperforms U-Net-based diffusion baselines (DiffusionMBIR, MCG), indicating strong domain sensitivity of their learned priors in this clinical OOD setting. Even advanced diffusion baselines such as DDS yield only marginal visual improvements, and severe sparse-view streaking remains visible in the reconstructions of all baselines (Row~2), including DDS and Restormer. In contrast, CDPIR substantially suppresses these artifacts while better preserving clinically relevant structures. The SSIM maps (Row~3) further highlight the improved structural agreement of CDPIR, with an SSIM gain of approximately $+0.09$ over ASD-POCS on this representative slice.

Quantitatively, Table~\ref{tab:clinical_data} confirms the same trend over the full clinical volume: compared with ASD-POCS, CDPIR increases PSNR from $28.58\pm0.42$ to $29.36\pm0.45$ and SSIM from $0.7853\pm0.0061$ to $0.8769\pm0.0049$, while substantially reducing LPIPS from $0.1874\pm0.0051$ to $0.0475\pm0.0072$. These results corroborate the qualitative comparison in Fig.~\ref{fig:ge} and demonstrate that CDPIR provides improved perceptual quality and structural fidelity under zero-shot clinical reconstruction with extreme view sparsity.

\subsection{Reconstruction on Extremity PCCT Dataset}
\begin{figure*}
    \centering
    \includegraphics[width=1\textwidth]{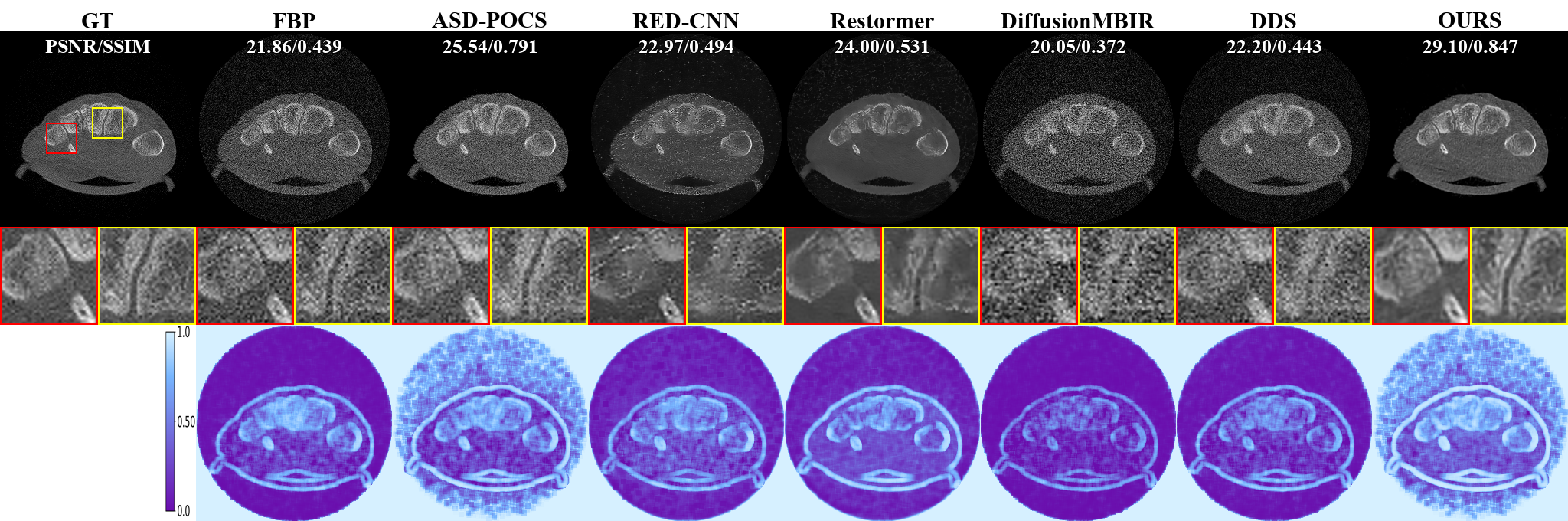}
    \caption{\textbf{Zero-shot reconstruction on MARS PCCT dataset.} For the 74-views CT reconstruction task, our model shows strong performance on the details of the image compared with other methods. The second row are magnified ROIs, and third row are the SSIM map between the prediction and the ground truth. The display window is [-700, 1200] HU}
\label{fig:pcct}
\end{figure*}

To enable fan-beam reconstruction, we rebin the raw 4D helical cone-beam projections into 2D fan-beam sinograms and reconstruct slice-by-slice. Each slice contains 373 views, from which we uniformly subsample 74 views to simulate a sparse-view setting. For evaluation, we reconstruct all slices from the $8^{\text{th}}$ energy bin. Figure~\ref{fig:pcct} shows a representative slice, and we compare CDPIR with several state-of-the-art diffusion-based and iterative baselines under a zero-shot setting (trained on AAPM only, without any pretraining or fine-tuning on PCCT).

As highlighted by the ROIs in the second row of Fig.~\ref{fig:pcct}, CDPIR effectively suppresses severe sparse-view artifacts while preserving fine anatomical structures in both soft tissue and bone. This PCCT dataset exhibits a substantial domain shift from AAPM (different scanner, acquisition physics, and noise characteristics), and can be considered a far-OOD reconstruction task~\cite{zhang2023openood}. Under this challenging setting, U-Net-based diffusion baselines generalize poorly and may underperform classical iterative reconstruction; for example, MCG fails to converge to a stable solution. In contrast, CDPIR yields markedly improved reconstruction fidelity.

Quantitatively, Table~\ref{tab:clinical_data} further confirms the robustness of CDPIR on PCCT: compared with ASD-POCS, CDPIR improves PSNR from $26.32\pm0.41$ to $29.60\pm0.91$ (+3.28 dB) and SSIM from $0.815\pm0.0051$ to $0.8659\pm0.0055$, while reducing LPIPS from $0.1221\pm0.0034$ to $0.1105\pm0.0099$. Together with Fig.~\ref{fig:pcct}, these results indicate that CDPIR better suppresses sparse-view artifacts and preserves fine anatomical detail in this highly distribution-shifted real-world dataset.

\begin{table*}[htbp]
\centering
\caption{ Quantitative Comparison of Clinical datasets including GE and PCCT datasets. Variances are shown in \textcolor{gray}{gray}.}
\label{tab:clinical_data}
\setlength{\tabcolsep}{3pt}
\renewcommand{\arraystretch}{1.3}
\begin{tabular}{l | ccc | ccc}
\toprule
\multirow{2}{*}{\textbf{Views}} &
\multicolumn{3}{c|}{\textbf{ASD-POCS}} &
\multicolumn{3}{c}{\textbf{CDPIR}} \\
\cmidrule(lr){2-4} \cmidrule(lr){5-7}
& \textbf{PSNR}$\uparrow$ & \textbf{SSIM}$\uparrow$ & \textbf{LPIPS}$\downarrow$
& \textbf{PSNR}$\uparrow$ & \textbf{SSIM}$\uparrow$ & \textbf{LPIPS}$\downarrow$ \\
\midrule
GE datasets &
28.58 \gvar{0.420} & 0.7853 \gvar{0.0061} & 0.1874 \gvar{0.0051} &\textbf{29.36}\gvar{0.447} & \textbf{0.8769} \gvar{0.0049} & \textbf{0.0475} \gvar{0.0072} \\
PCCT datasets& 
26.32\gvar{0.41} & 0.815 \gvar{0.0051} & 0.1221\gvar{0.0034} &\textbf{29.60}\gvar{0.91} & \textbf{0.8659} \gvar{0.0055} & \textbf{0.1105}\gvar{0.0099}  \\
\bottomrule
\end{tabular}
\end{table*}

\subsection{Reconstruction on Multiple View Settings}
In our simulation experiments, we use 55 projection views as the default sparse-view setting for CT reconstruction. To further evaluate the robustness of CDPIR under different levels of data sparsity, we conduct additional experiments with 35, 55, 65, and 75 views. In all cases, the diffusion prior is pretrained on the AAPM dataset and reconstruction is performed on the XCAT dataset, constituting an OOD evaluation scenario. This multi-view protocol enables a systematic assessment of CDPIR as the number of available projections increases, covering regimes from severely undersampled to moderately sparse. Quantitative results are reported in Table~\ref{tab:multiview}.

\begin{table*}[htbp]
\centering
\caption{ Quantitative comparison under multiple view settings. Standard deviations are shown in \textcolor{gray}{gray}.}
\label{tab:multiview}
\setlength{\tabcolsep}{3pt}
\renewcommand{\arraystretch}{1.3}

\begin{tabular}{l | ccc | ccc | ccc}
\toprule
\multirow{2}{*}{\textbf{Views}} & \multicolumn{3}{c|}{\textbf{DDS}} & \multicolumn{3}{c|}{\textbf{DDIP3D}} & \multicolumn{3}{c}{\textbf{CDPIR}} \\
\cmidrule(lr){2-4} \cmidrule(lr){5-7} \cmidrule(lr){8-10}
& \textbf{PSNR}$\uparrow$ & \textbf{SSIM}$\uparrow$ & \textbf{LPIPS}$\downarrow$
& \textbf{PSNR}$\uparrow$ & \textbf{SSIM}$\uparrow$ & \textbf{LPIPS}$\downarrow$
& \textbf{PSNR}$\uparrow$ & \textbf{SSIM}$\uparrow$ & \textbf{LPIPS}$\downarrow$ \\
\midrule

35&
30.07\gvar{1.16} & 0.8074\gvar{0.0103} & 0.134\gvar{0.0081} &
34.64\gvar{1.50} & 0.856\gvar{0.0205} & 0.0205\gvar{0.0034} &
\textbf{36.97}\gvar{0.92} & \textbf{0.9545}\gvar{0.0042} & \textbf{0.0207}\gvar{0.0027} \\

55&
32.22\gvar{1.98} & 0.881\gvar{0.0006} & 0.028\gvar{0.0800} &
34.78\gvar{4.46} & 0.908\gvar{0.0004} & 0.021\gvar{0.0000} &
\textbf{40.19}\gvar{1.59} & \textbf{0.9631}\gvar{0.0078} & \textbf{0.0185}\gvar{0.0059} \\

65&
34.52\gvar{1.48} & 0.9046\gvar{0.0071} & 0.0714\gvar{0.0055} &
37.90\gvar{1.53} & 0.9036\gvar{0.0123} & 0.0081\gvar{0.0008} &
\textbf{43.73}\gvar{1.11} & \textbf{0.9837}\gvar{0.0009} & \textbf{0.0069}\gvar{0.0008} \\

75&
34.88\gvar{2.41} & 0.9231\gvar{0.0056} & 0.0566\gvar{0.0046} &
38.02\gvar{1.66} & 0.9044\gvar{0.0152} & 0.0068\gvar{0.0007} &
\textbf{43.95}\gvar{0.54} & \textbf{0.9820}\gvar{0.0010} & \textbf{0.0065}\gvar{0.0008} \\

\bottomrule
\end{tabular}
\end{table*}

\subsection{2D vs.\ 3D Diffusion Analysis}
 Because CDPIR relies on a 2D slice-wise diffusion prior, a natural concern is
whether stochastic sampling introduces inter-slice discontinuities in the
reconstructed volume.  To address this, we conduct three complementary
validations on an ID volumetric reconstruction, consisting of 57 consecutive
axial slices from the COCA dataset.

\subsubsection{Quantitative Inter-Slice Consistency}
\begin{figure}[htbp]
  \centering
  \includegraphics[width=0.95\linewidth]{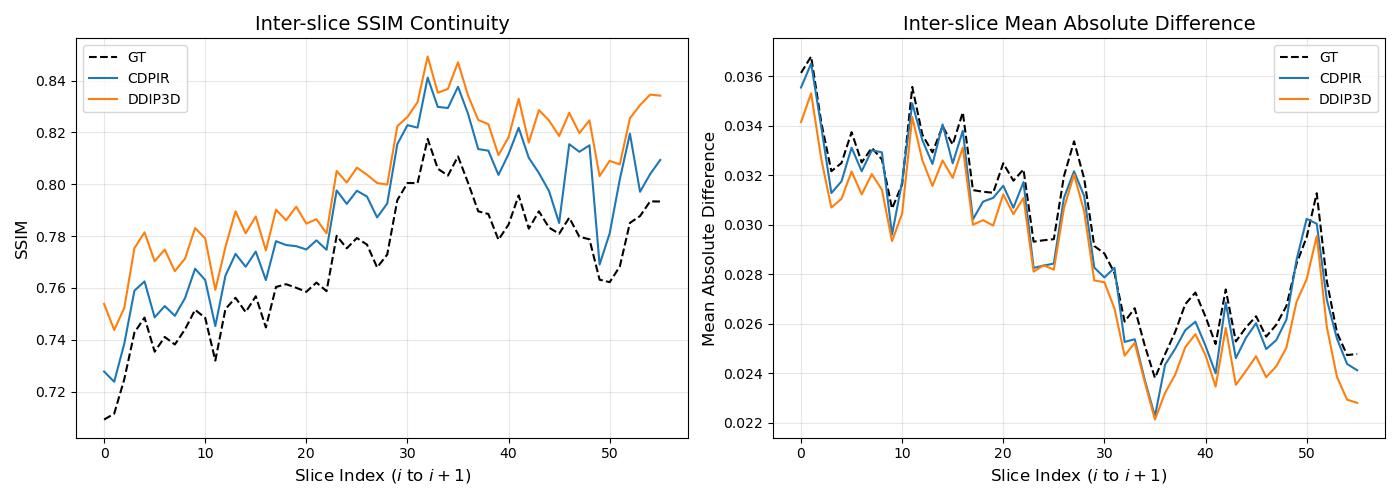}
  \caption{%
    Inter-slice consistency metrics computed on 57 consecutive slices
    (COCA dataset, ID).
    \textbf{Left:} Inter-slice $SSIM (x_i,x_{i+1})$.
    \textbf{Right:} Masked inter-slice Mean Absolute Difference
    $\|x_i - x_{i+1}\|_1$.
    }
  \label{fig:interslice_metrics}
\end{figure}

We reconstruct the full 3D volume and compute two complementary metrics for
every adjacent slice pair $(x_i, x_{i+1})$:
(i)~inter-slice SSIM, $\mathrm{SSIM}(x_i, x_{i+1})$, which measures structural
coherence between consecutive slices, and
(ii)~masked inter-slice Mean Absolute Difference (MAD),
$\|x_i - x_{i+1}\|_1$, computed within the body mask.
Fig.~\ref{fig:interslice_metrics} plots curves for CDPIR, DDIP3D, and the
ground truth (GT) along the slice index.  
\red{Compared with DDIP3D, CDPIR's curves more closely follow the GT trend and
remain smooth throughout the image volume, without abrupt spikes or oscillations that
would indicate visually disruptive slice-to-slice discontinuities.
We emphasize that inter-slice SSIM and MAD are descriptive measures of
slice-to-slice smoothness, not direct measures of volumetric fidelity. In
particular, inter-slice SSIM values higher than those of the GT reference should
not be interpreted as better anatomical continuity; rather, they may indicate
reduced through-plane variation and potential z-axis over-smoothing, which is a
possible limitation of slice-wise 2D reconstruction. Accordingly, we interpret
Fig.~\ref{fig:interslice_metrics} only as evidence that CDPIR maintains stable
slice transitions without severe discontinuities.}

\subsubsection{Qualitative Multi-Planar Reformat Visualization}
\begin{figure}[htbp]
  \centering
  \includegraphics[width=0.95\linewidth]{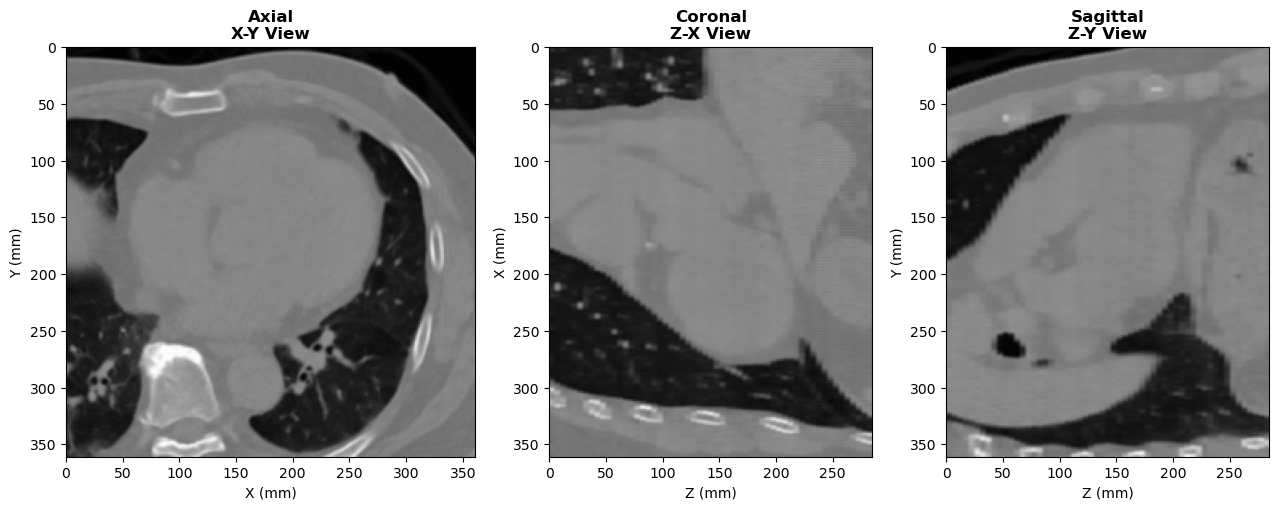}
  \caption{%
    Multi-planar reformats of a representative CDPIR reconstruction
    (COCA dataset): axial X–Y view (left), coronal Z–X view (centre),
    and sagittal Z–Y view (right).  Continuous anatomy is preserved
    across the volume without visible slice gaps or banding artifacts.}
  \label{fig:mpr}
\end{figure}

To provide a holistic view of volumetric coherence, Fig.~\ref{fig:mpr} shows
multi-planar reformats (MPR) of the CDPIR reconstruction in the axial (X–Y),
coronal (Z–X), and sagittal (Z–Y) planes.  Continuous anatomical structures
such as the cardiac chambers, ribs, and soft-tissue boundaries appear smooth and
uninterrupted across all three orthogonal views, confirming the absence of
slice-boundary artifacts or discontinuities that would otherwise manifest as
horizontal banding in the coronal and sagittal reformats.

\subsubsection{Comparison with a 2.5D Baseline as a Proxy for 3D Context}
Training a full 3D diffusion model is substantially more memory- and compute-intensive in our current setting.
As a practical proxy for explicit 3D context, we implement a 2.5D diffusion baseline, DDIP3D-2.5D, following its standard inference design:
each forward path takes \textbf{7 adjacent slices} as input and predicts the \textbf{central 5 slices}.
To reconstruct the full 57-slice volume, we apply DDIP3D-2.5D in a sliding-window fashion and stitch the predicted central slices across windows, so that all methods are evaluated on the same set of output slices.
For completeness, we also implement a 2.5D variant of our method (CDPIR-2.5D).
In our implementation, CDPIR-2.5D conditions on the \textbf{entire 57-slice volume} as contextual input when reconstructing each slice.
This setting is used only as an \emph{oracle upper bound} on the potential benefit of additional inter-slice context, rather than as a fairness-critical baseline.
Notably, CDPIR(2D) remains competitive with CDPIR(2.5D), suggesting limited incremental gain from explicit 2.5D conditioning under our DC-coupled solver.
All models are evaluated on the same ID COCA volume, and we report PSNR/SSIM/LPIPS measured on axial, coronal, and sagittal multi-planar reformats (MPR).

Table~\ref{tab:comparison_3d_all_views} summarizes the results.
First, \textbf{CDPIR(2D) consistently outperforms DDIP3D-2.5D} on every metric and every view, with statistically significant margins (two-sided paired $t$-test; $p \ll 0.001$ for PSNR/SSIM/LPIPS in all views).
The PSNR gains range from $+1.99$\,dB (axial) to $+4.65$\,dB (sagittal), SSIM improves by up to $+0.023$, and LPIPS decreases by up to $-0.006$.
These results indicate that simply injecting local inter-slice context via a 2.5D input does not close the gap to CDPIR.
Second, comparing \textbf{CDPIR(2D) vs.\ CDPIR(2.5D)}, we observe that the performance is broadly similar across all views, with only small differences.
Particularly, CDPIR(2D) achieves slightly higher PSNR on all three views, while SSIM/LPIPS are close between the two variants.
This suggests that the iterative data-consistency coupling in CDPIR already provides strong volumetric regularization, and explicit 2.5D conditioning offers limited additional benefit under our setting.

\begin{table}[htbp]
\centering
\caption{Comparison of DDIP3D(2.5D) and CDPIR with 2D/2.5D priors on COCA
  (mean $\pm$ std). $^{*}p<0.001$ denotes CDPIR(2D) vs.\ DDIP3D(2.5D), paired $t$-test.}
\label{tab:comparison_3d_all_views}
\resizebox{\columnwidth}{!}{%
\begin{tabular}{@{}llccc@{}}
\toprule
View & Metric & DDIP3D (2.5D) & CDPIR (2.5D) & CDPIR (2D) \\
\midrule
\multirow{3}{*}{\textbf{Axial}}
  & PSNR\,$\uparrow$    & $39.28\pm0.54$ & $40.85\pm0.75$ & $\mathbf{41.27\pm0.92}^{*}$ \\
  & SSIM\,$\uparrow$    & $0.9509\pm0.0035$ & $\mathbf{0.9683\pm0.0117}$ & $0.9676\pm0.0114^{*}$ \\
  & LPIPS\,$\downarrow$ & $0.0327\pm0.0073$ & $0.0271\pm0.0156$ & $\mathbf{0.0265\pm0.0152}^{*}$ \\
\midrule
\multirow{3}{*}{\textbf{Coronal}}
  & PSNR\,$\uparrow$    & $35.22\pm2.69$ & $41.35\pm0.94$ & $\mathbf{42.04\pm0.98}^{*}$ \\
  & SSIM\,$\uparrow$    & $0.9423\pm0.0219$ & $\mathbf{0.9787\pm0.0056}$ & $0.9766\pm0.0061^{*}$ \\
  & LPIPS\,$\downarrow$ & $0.0288\pm0.0080$ & $\mathbf{0.0174\pm0.0084}$ & $\mathbf{0.0174\pm0.0082}^{*}$ \\
\midrule
\multirow{3}{*}{\textbf{Sagittal}}
  & PSNR\,$\uparrow$    & $35.56\pm2.61$ & $41.52\pm0.99$ & $\mathbf{42.16\pm0.86}^{*}$ \\
  & SSIM\,$\uparrow$    & $0.9451\pm0.0064$ & $\mathbf{0.9808\pm0.0033}$ & $0.9790\pm0.0028^{*}$ \\
  & LPIPS\,$\downarrow$ & $0.0269\pm0.0049$ & $\mathbf{0.0192\pm0.0076}$ & $0.0195\pm0.0076^{*}$ \\
\bottomrule
\end{tabular}}
\vspace{0.35em}

\end{table}

Taken together with the slice-consistency metrics and the MPR visualizations, this comparative study supports that CDPIR's 2D diffusion prior does not introduce noticeable inter-slice variation and maintains strong volumetric reconstruction quality across anatomical views.

\section{Ablation Study}

To quantify the contribution of each component in the model, we conduct ablation experiments along three axes: (i) backbone architecture (CNN vs.\ Transformer), (ii) model scaling (capacity), and (iii) sampling behavior (guidance/conditioning and generative transport). Unless otherwise noted, all ablations use the same training data splits, the same data-consistency module (ASD-POCS), and identical inference hyperparameters.

\subsection{Architecture Isolation: U-Net vs. Transformer}
\begin{table}[htbp]
    \centering
    \caption{Quantitative comparison of backbone architectures evaluated on the OOD XCAT dataset (55 views). Both models were trained on the combined AAPM and COCA datasets.}
    \label{tab:backbone}
    \resizebox{0.48\textwidth}{!}{ 
        \begin{tabular}{lccccc}
            \toprule
            \textbf{Method} & \textbf{Backbone} &\textbf{Model capacity} & \textbf{PSNR}$\uparrow$ & \textbf{SSIM}$\uparrow$ & \textbf{LPIPS}$\downarrow$ \\
            \midrule
            U-Net Pipeline & Guided-Diffusion & 374.0M & 35.68 $\pm$ 1.37 & 0.919 $\pm$ 0.0049 & 0.0535 $\pm$ 0.0041 \\
            \textbf{CDPIR (Ours)} & \textbf{SiT-B} & 142.9M & \textbf{38.36 $\pm$ 0.75} & \textbf{0.953 $\pm$ 0.0028} & \textbf{0.0180 $\pm$ 0.0026} \\
            \bottomrule
        \end{tabular}
    }
\end{table}

Based on our key experiments, one can see that the transformer backbone significantly improves OOD reconstruction quality compared to U-Net-based architectures. 
To explicitly evaluate the contribution of the transformer-based backbone compared to traditional CNN-based architectures, we perform a rigorous controlled experiment. We construct a strong CNN baseline based on the widely used Guided-Diffusion U-Net architecture. To ensure a fair comparison, we align the training and inference strategies of this U-Net baseline strictly with our CDPIR framework. We adapt the guided-diffusion for CFG instead of using an external classifier guidance. And the guided-diffusion baseline is integrated into the same ASD-POCS iterative framework. Both models are trained on the combined AAPM and COCA datasets and evaluated on the XCAT dataset to assess OOD generalizability. 

As shown in Table \ref{tab:backbone}, under the same training and physical constraint conditions, the Transformer-based CDPIR significantly outperforms the U-Net baseline on the OOD 55-views XCAT task. While the U-Net baseline benefits from the CFG and ASD-POCS integration, it lags behind CDPIR by $2.68$ dB in PSNR and  $0.034$ in SSIM.

\begin{figure}[t]
    \centering
    \includegraphics[width=0.48\textwidth]{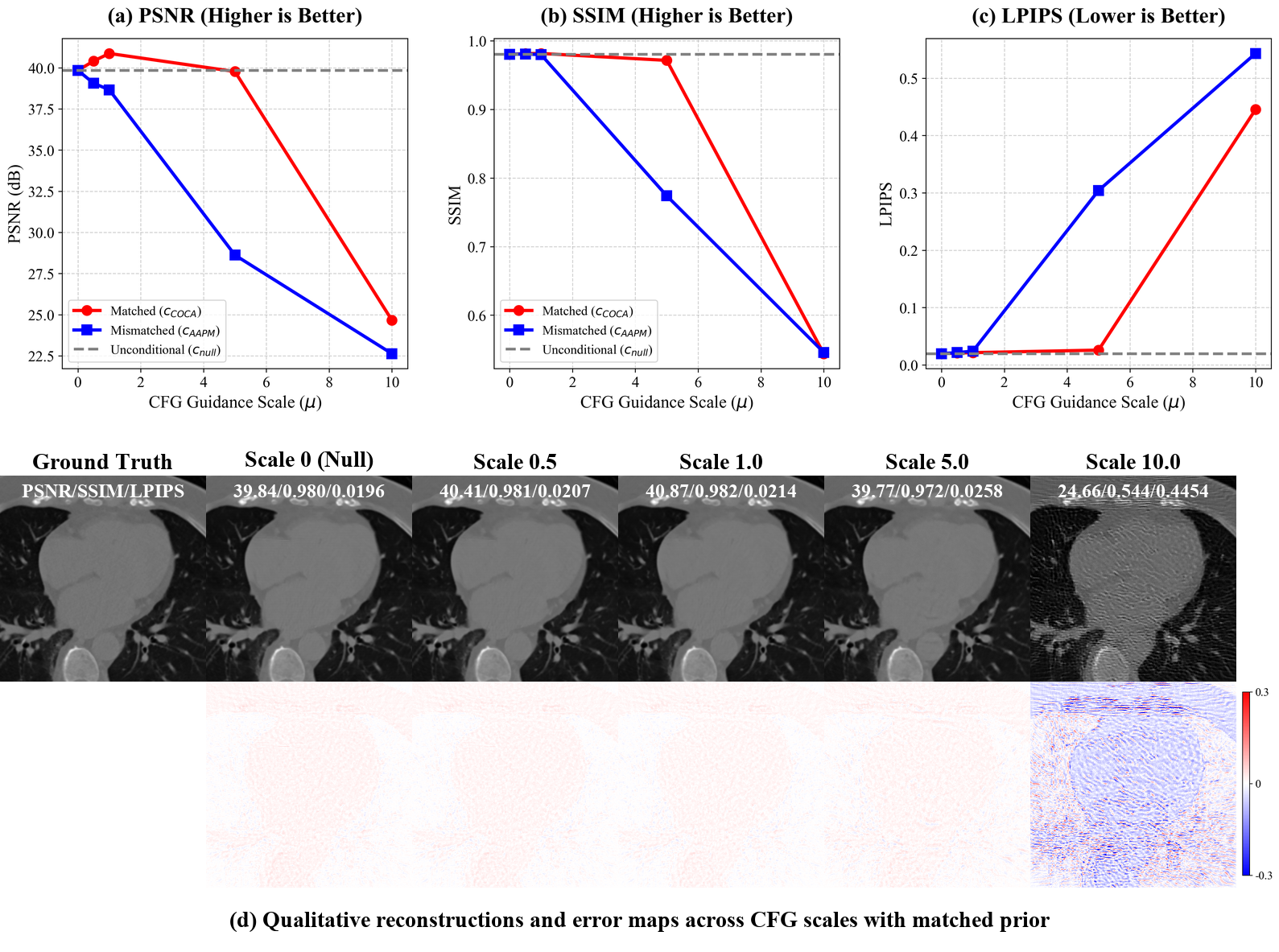}
    \caption{\textbf{Effect of CFG scale and conditional input during sampling.} \textbf{(a--c)} Quantitative performance (PSNR/SSIM/LPIPS) as a function of CFG guidance scale $\mu$ under different conditioning: matched prior ($c_{\mathrm{COCA}}$, red), mismatched prior ($c_{\mathrm{AAPM}}$, blue), and unconditional ($c_{\mathrm{null}}$, gray dashed). \textbf{(d)} Qualitative reconstructions and corresponding error maps under the matched prior across different $\mu$. Moderate guidance (e.g., $\mu\!\approx\!0.5\!-\!1$) improves reconstruction quality, while overly large $\mu$ destabilizes sampling and degrades fidelity/perceptual quality.}
    \label{fig:cfg_scale}
\end{figure}

\subsection{Effect of CFG Scale and Conditional Input}

To investigate the role of CFG at inference time, we perform an inference experiment on a representative case of the COCA dataset. We  vary the guidance scale $\mu$ while keeping all other settings fixed. We evaluate $\mu \in \{0, 0.5, 1.0, 5.0, 10.0\}$ and compare three conditioning choices during sampling: the AAPM token ($c_{\mathrm{AAPM}}$, class 0), the COCA token ($c_{\mathrm{COCA}}$, class 1), and the unconditional branch ($c_{\mathrm{null}}$).
Following Eq.~\eqref{v_cal}, the CFG scale $\mu$ balances the null-conditioned term and the conditional residual. Particularly, when $\mu=0$, the update reduces exactly to the unconditional branch regardless of the provided label, so different conditioning inputs become equivalent in this limit. As $\mu$ increases, the conditional residual increasingly dominates, which can amplify domain-specific bias and destabilize sampling under distribution shift if the conditioning token is mismatched.

Fig.~\ref{fig:cfg_scale} shows that moderate guidance ($\mu \approx 0.5\text{--}1.0$) improves OOD reconstruction when using a \emph{matched} conditioning token (here $c_{\mathrm{COCA}}$), yielding the best PSNR/SSIM while maintaining low LPIPS. \red{In contrast, using a \emph{mismatched} token (here $c_{\mathrm{AAPM}}$) becomes increasingly harmful as $\mu$ grows, consistent with our interpretation that the conditional residual encodes domain-specific appearance statistics that may conflict with OOD measurements.} Finally, overly large guidance (\emph{e.g.}, $\mu=10$) leads to severe structured artifacts and a collapse in fidelity and perceptual quality, as also evidenced by the qualitative reconstructions and error maps in Fig.~\ref{fig:cfg_scale}(d). Based on this study, we adopt $\mu=1$ and the matched conditioning token for all subsequent OOD experiments unless otherwise stated. For OOD data without a dedicated token, a practical strategy is to evaluate the available tokens and select the one that yields the smallest data-fidelity residual $\|Ax-y\|_2$ after a few outer iterations; the unconditional setting ($\mu=0$) serves as a robust fallback.

\subsection{Evaluation of the Cross-Distribution Prior}
CDPIR is designed to learn a cross-distribution generative prior from multiple datasets.
During training, we adopt CFG by randomly dropping the dataset/domain token with a fixed probability and replacing it with a null embedding.
This encourages the model to learn both (i)  an unconditional branch that emphasizes anatomy-level structure shared across domains and (ii) a conditional branch that emphasizes domain-specific characteristics, which is important for robustness under distribution shifts.
To verify the necessity of CFG, we perform an ablation where CFG is removed.
Specifically, we train the model using all datasets but force a single shared token (class label $c{=}0$) for every sample and disable the conditional-drop mechanism.
All other settings (network architecture, training data, and inference pipeline including the same data consistency module) are kept unchanged.

Table~\ref{tab:ablation_cfg} reports the results on the AAPM$\rightarrow$XCAT setting (35 and 55 views).
Training without CFG consistently degrades reconstruction accuracy, reducing PSNR/SSIM and increasing LPIPS.
These results show that the conditional-drop training strategy contributes to learning a more
transferable cross-distribution prior and improves OOD reconstruction
performance. \red{More importantly, they are consistent with our interpretation that
CFG encourages a complementary decomposition between a shared structural prior
and a domain-specific residual modulation.} Overall, CFG improves
reconstruction fidelity across cases, especially under severe undersampling.

\begin{table*}[htbp]
    \centering
    \caption{Ablation on classifier-free guidance (CFG) training evaluated on the OOD XCAT dataset under 35 and 55 views. All settings are identical except that ``w/o CFG'' disables conditional-drop and uses a single shared domain token for all training samples.}
    \resizebox{\linewidth}{!}{
    \begin{tabular}{lccc ccc}
        \toprule
        & \multicolumn{3}{c}{\textbf{35 views}} & \multicolumn{3}{c}{\textbf{55 views}} \\
        \cmidrule(lr){2-4}\cmidrule(lr){5-7}
        \textbf{Training Setting} & \textbf{PSNR}$\uparrow$ & \textbf{SSIM}$\uparrow$ & \textbf{LPIPS}$\downarrow$
                                 & \textbf{PSNR}$\uparrow$ & \textbf{SSIM}$\uparrow$ & \textbf{LPIPS}$\downarrow$ \\
        \midrule
        w/o CFG (single token, no drop) 
            & 36.91 $\pm$ 0.067 & 0.9486 $\pm$ 0.0000 & 0.0247 $\pm$ 0.0000
            & 41.16 $\pm$ 1.09  & 0.9768 $\pm$ 0.0014 & 0.0099 $\pm$ 0.0010 \\
        \textbf{w/ CFG (Ours)} 
            & \textbf{36.97 $\pm$ 0.092} & \textbf{0.9545 $\pm$ 0.0042} & \textbf{0.0207 $\pm$ 0.0027}
            & \textbf{41.74 $\pm$ 1.69} & \textbf{0.9797 $\pm$ 0.0012} & \textbf{0.0088 $\pm$ 0.0010} \\
        \bottomrule
    \end{tabular}
    }
    \label{tab:ablation_cfg}
\end{table*}


\subsection{Impact of Generative Framework}
To evaluate the effect of the stochastic interpolant, we replace SiT in CDPIR
with a Variance Preserving Stochastic Differential Equation (VP-SDE) diffusion
formulation (DiT-style) while keeping the rest of the pipeline unchanged.
Specifically, both methods use the same Transformer backbone, are trained on the
same dataset, and share the same inference components, including the same
classifier-free guidance strategy and the same ASD-POCS data-consistency (DC)
operator applied at each sampling step. Therefore, the performance gap can be
attributed primarily to the generative dynamics (velocity/interpolant vs.\
score/VP-SDE), rather than to sampling or DC settings.

Therefore, the performance gap can be attributed primarily to the generative dynamics (velocity/interpolant vs.\ score/VP-SDE), rather than sampling or DC settings.

\begin{figure}[t]
\centering
\includegraphics[width=0.49\textwidth]{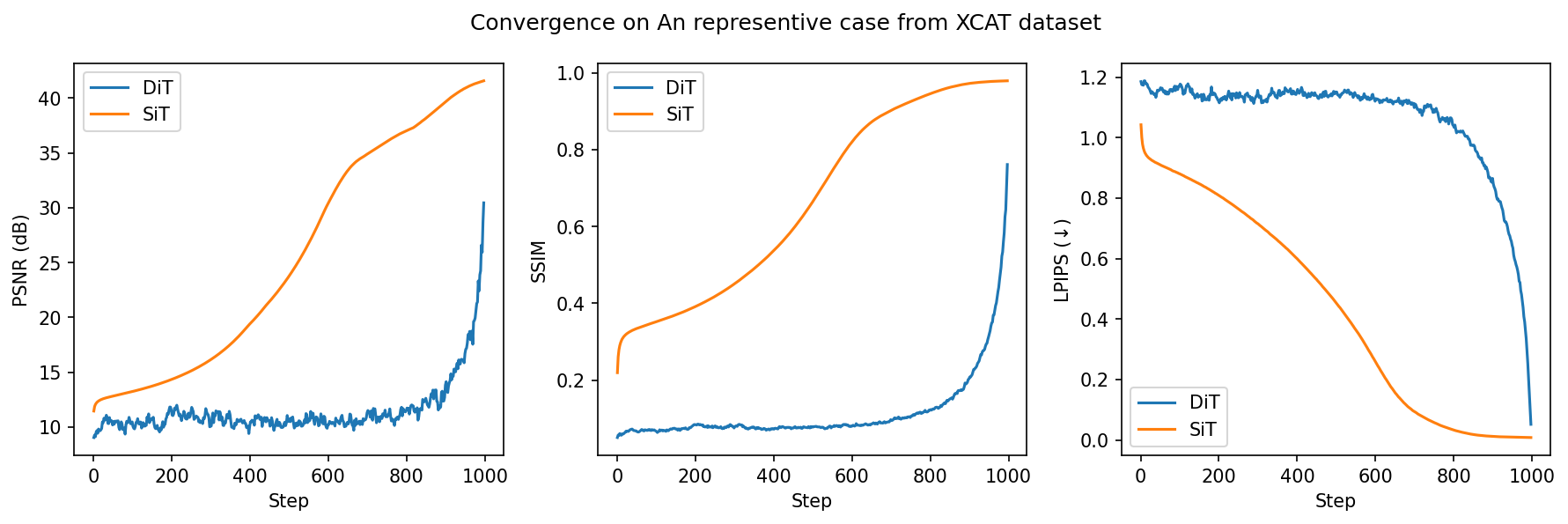} 
\caption{\textbf{Step-wise convergence of reconstruction quality.} PSNR/SSIM/LPIPS versus sampling step ($N{=}1,000$) for SiT (stochastic interpolant) and the VP-SDE baseline.}
\label{fig:ablation_convergence}
\end{figure}

Fig.~\ref{fig:ablation_convergence} compares the step-wise convergence of reconstruction quality over $N{=}1,000$ sampling steps on a representative XCAT case.
SiT exhibits a smooth and monotonic recovery in PSNR/SSIM and a stable decrease in LPIPS, indicating efficient transport from noise to the data manifold.
In contrast, the VP-SDE baseline stays in a low-SNR regime for most of the trajectory and improves sharply only near the end, accompanied by higher perceptual volatility.

\begin{figure}[t]
\centering
\includegraphics[width=0.49\textwidth]{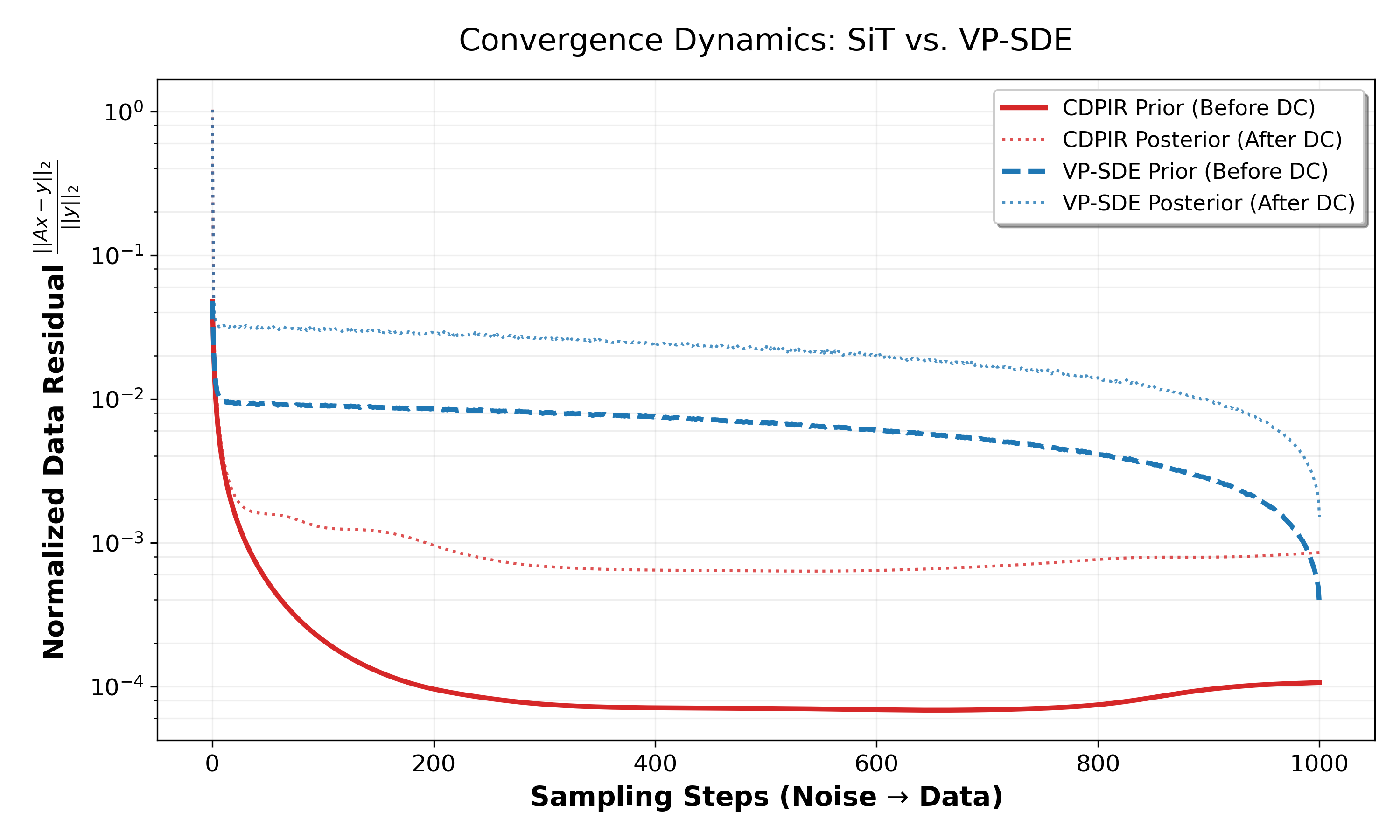} 
\caption{\textbf{Data-fidelity residual along the trajectory.} Normalized residual $\|Ax-y\|/\|y\|$ at each sampling step ($N{=}1,000$), reported before DC (prior) and after DC (posterior).}
\label{fig:dc_residual}
\end{figure}

To further examine measurement consistency, Fig.~\ref{fig:dc_residual} reports the normalized data-fidelity residual $\|Ax-y\|/\|y\|$ before DC (prior) and after DC (posterior) at each sampling step.
SiT yields faster decay of the post-DC residual and reaches a substantially lower residual floor, suggesting that interpolant-driven intermediate states are more DC-friendly and allow ASD-POCS to enforce consistency more effectively throughout the trajectory.
Overall, the two figures consistently demonstrate that the stochastic interpolant improves both convergence efficiency and compatibility with iterative data consistency under the same inference settings.

\begin{itemize}
\item \textbf{Trajectory efficiency (quality metrics).}
SiT shows a smooth and monotonic improvement in PSNR/SSIM with a stable LPIPS decrease (Fig.~\ref{fig:ablation_convergence}), indicating efficient transport from noise to the data manifold.
VP-SDE stays in a low-SNR regime for most steps and improves only near the end, with higher perceptual volatility.

\item \textbf{Compatibility with data consistency (DC residual).}
Under the same ASD-POCS DC budget, SiT yields faster decay of the post-DC residual and reaches a lower residual floor (Fig.~\ref{fig:dc_residual}),
suggesting that interpolant-driven intermediate states are more DC-friendly and enable more effective enforcement of measurement consistency.
\end{itemize}


 \subsection{Impact of Model Capacity}

\begin{figure}[t]
    \centering
    \includegraphics[width=0.75\linewidth]{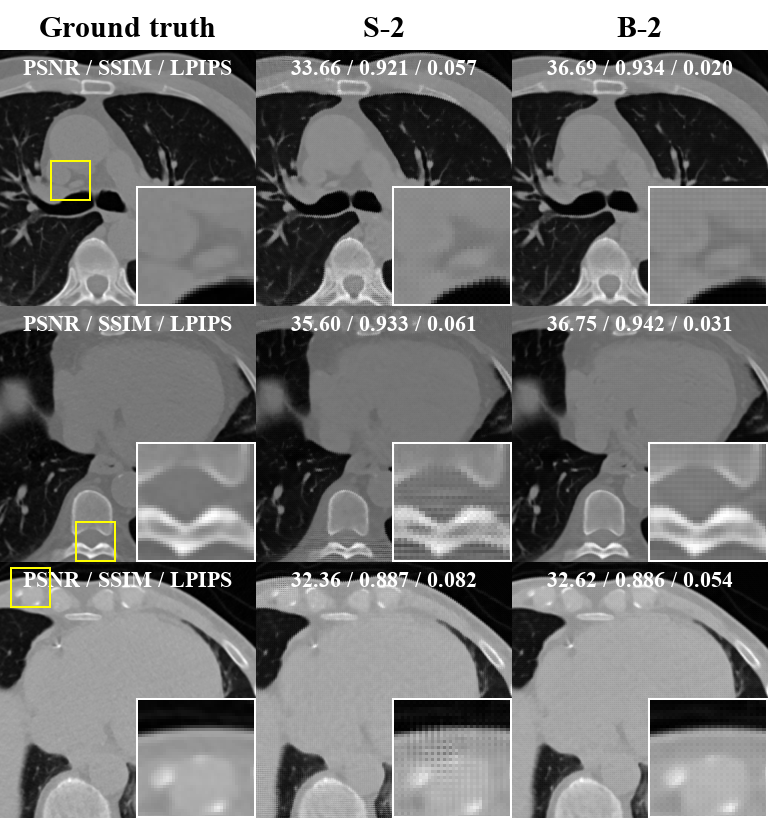}
    \caption{Comparison of S-2 and B-2 reconstructions on AAPM-COCA. Yellow boxes mark ROIs; bottom-right insets show magnified views. Display window: [-1000,1000] HU.}
    \label{fig:model_ablation}
\end{figure}

\begin{table}[t]
\scriptsize
\setlength{\tabcolsep}{3.5pt}
\centering
\caption{OOD performance (PSNR/SSIM/LPIPS, mean$\pm$std) for small (S-2) and big (B-2) models trained on AAPM.}
\label{tab:model_scaling_ablation}
\resizebox{0.48\textwidth}{!}{
\begin{tabular}{cc|ccc|ccc}
\toprule
$q(x)$ & $\bar{q}(x)$
& \multicolumn{3}{c|}{\textbf{S-2}}
& \multicolumn{3}{c}{\textbf{B-2}} \\
\cmidrule(lr){3-5}\cmidrule(lr){6-8}
& & PSNR$\uparrow$ & SSIM$\uparrow$ & LPIPS$\downarrow$
  & PSNR$\uparrow$ & SSIM$\uparrow$ & LPIPS$\downarrow$ \\
\midrule
AAPM & COCA
& 35.54$\pm$2.14 & 0.9218$\pm$0.0222 & 0.0628$\pm$0.0161
& 40.42$\pm$3.23 & 0.963$\pm$0.0264 & 0.037$\pm$0.0034 \\
AAPM & XCAT
& 35.78$\pm$1.42 & 0.9445$\pm$0.0027 & 0.0276$\pm$0.0030
& 38.36$\pm$0.75 & 0.953$\pm$0.0028 & 0.018$\pm$0.0026 \\
\bottomrule
\end{tabular}
}
\end{table}
Transformer-based diffusion models enable flexible scaling through architectural variations. The model capacities follow the scaling scheme of SiT~\cite{ma2024sit}, with CDPIR-B-2 adopting the DiT-B parameterization and CDPIR-S-2 its smaller variant. Their specifications are summarized in Table~\ref{tab:model_size}.
Trained on AAPM and tested on COCA and XCAT, B-2 consistently outperforms S-2 (Table~\ref{tab:model_scaling_ablation}), with gains of up to +4 dB PSNR and +0.03 SSIM on COCA and smaller but steady improvements on XCAT. Fig.~\ref{fig:model_ablation} further shows sharper edges, clearer bone structures, and more accurate coronary arteries in B-2 reconstructions. These results demonstrate that scaling model capacity enhances both reconstruction accuracy and structural fidelity, highlighting the importance of architectural scalability for OOD robustness in transformer-based diffusion models.

Beyond the qualitative observations in Fig.~\ref{fig:model_ablation}, we find that the blocky patterns are spatially aligned with the $p=2$ patch boundaries and remain consistent across different random seeds, suggesting a token-lattice imprint rather than stochastic sampling noise. This behavior is amplified when operating directly in image space without an autoencoder, since the model must simultaneously model fine-scale texture and enforce smooth cross-patch transitions.

These observations empirically validate that the performance gains of CDPIR are not merely due to the Transformer backbone, but stem from the superior transport properties of the stochastic interpolant framework. By constructing a straighter and more deterministic path between the prior and data distributions, our method achieves higher reconstruction fidelity and stability under the same computational budgets.

\section{Discussion}
Our experiments demonstrate that CDPIR remains robust under challenging distribution shifts and can effectively handle OOD reconstruction scenarios. Compared with traditional iterative methods and the state-of-the-art diffusion baseline DDS~\cite{chung2023decomposed}, CDPIR better preserves fine anatomical structures while suppressing streaking and other sampling-induced artifacts, yielding more faithful reconstructions in both in-distribution and OOD settings.

\begin{figure}[t] 
\centering 
\includegraphics[width=\linewidth]{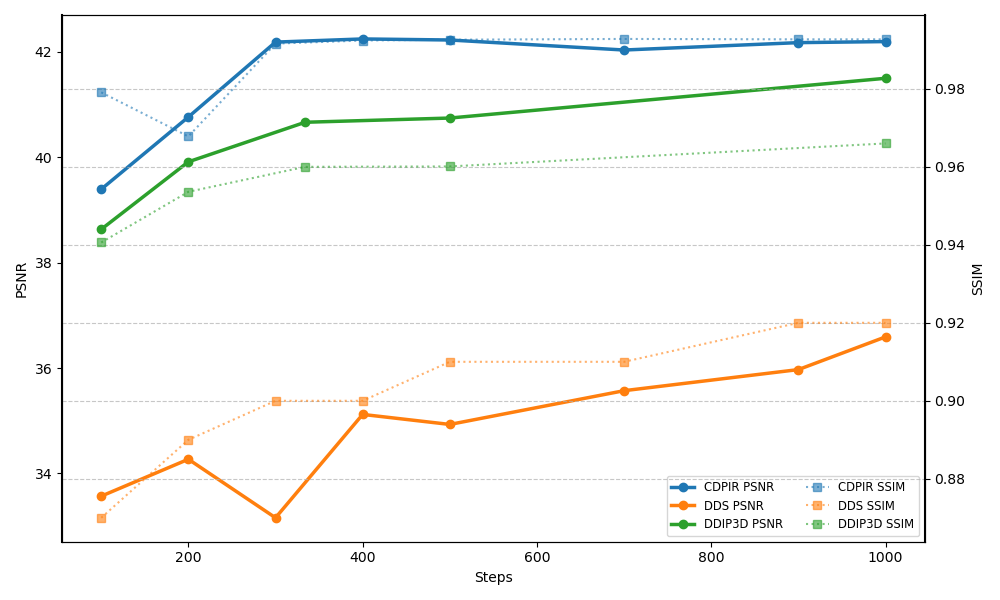} \caption{Quantitative comparison between DDS, DDIP3D, and CDPIR with respect to different sampling steps (100 to 1,000) on a representative AAPM-COCA testing case. CDPIR maintains superior quality with fewer steps.} 
\label{fig:steps_ablation} 
\end{figure} 

To evaluate practical convergence and sampling efficiency, we evaluate reconstruction quality with a reduced number of sampling steps from 100 to 1,000 using a representative OOD testing case from the AAPM--COCA setting. We compare DDS~\cite{chung2023decomposed}, DDIP3D~\cite{chung2024deep}, and CDPIR. As shown in Fig.~\ref{fig:steps_ablation}, DDS and DDIP3D typically require close to 1,000 steps to reach acceptable quality, whereas CDPIR achieves high-quality reconstructions with substantially fewer steps. This behavior is consistent with the improved numerical stability of the interpolant-based stochastic process used in our SiT formulation, where the velocity parameterization yields smoother and more stable convergence dynamics. In practice, we find that 300 steps already provide satisfactory reconstruction quality while significantly reducing inference time.

 Despite these gains, diffusion-based reconstruction remains inherently iterative and therefore often requires higher runtime than one-shot regressors. Table~\ref{tab:runtime} summarizes the average inference time per slice under 1,000 sampling steps, peak GPU memory footprint, and parameter count for representative diffusion-based baselines and CDPIR. We also report CDPIR-300 with 300 sampling steps. With 300 steps, CDPIR reconstructs one full CT scan in 507.2\,s (8.45\,min) with 2{,}194\,MB peak GPU memory and 142.9\,M parameters, whereas DDIP3D requires 23.31\,min with 8{,}252\,MB peak memory and 374.0\,M parameters. While the current runtime may still be high for time-critical clinical workflows, several practical strategies can further reduce computation without changing the reconstruction objective: reduced-step schedules and early stopping enabled by the fast convergence observed in Fig.~\ref{fig:steps_ablation}; deterministic probability-flow ODE sampling to eliminate stochastic overhead; mixed-precision inference and kernel/operator fusion for Transformer blocks; and more efficient projector/backprojector implementations, including caching and reuse across iterations. These optimizations are orthogonal to CDPIR and can be incorporated to improve throughput and facilitate near-real-time deployment.

\begin{table}[t] 
\centering 
\caption{Average inference time per slice, peak GPU memory usage, and number of learnable parameters. All methods are evaluated under their default inference configurations on the same hardware. Times are reported in minutes, with the corresponding mm:ss shown in parentheses.} 
\label{tab:runtime} 
\resizebox{0.48\textwidth}{!}{ 
\begin{tabular}{lccc} 
\toprule Method & Time / slice (min) & Peak GPU mem. (MB) & Params (M) \\ 
\midrule DiffusionMBIR & 1.37 (1:22) & 2884 & 61.4 \\ DDS & 5.25 (5:15) & 7333 & 374.0 \\ DDIP3D & 23.31 (23:19) & 8252 & 374.0 \\ CDPIR (Ours) & 22.83 (22:50) & 2194 & 142.9 \\ CDPIR-300 & 8.45 (8:27) & 2194 & 142.9 \\ 
\bottomrule 
\end{tabular} } 
\end{table}

The strong generalizability of CDPIR is partly attributable to its Transformer-based architecture. Compared with U-Net diffusion models that may be constrained by CNN inductive biases, Transformers capture long-range dependencies, scale effectively with model capacity, and propagate correction signals across regions with heterogeneous reliability. This design not only improves reconstruction fidelity but also provides a flexible foundation for exploring alternative inductive biases, such as state-space models like Mamba~\cite{gu2023mamba}, which introduce stronger sequential structure and may further improve efficiency in structured volumetric data.

We also observe an asymmetric transfer effect in which training on \textbf{AAPM} can sometimes match or outperform training on \textbf{COCA} even when testing on \textbf{COCA}, as reflected by the AAPM$\rightarrow$COCA direction in Fig.~\ref{fig:ood_heatmaps}. This can be explained by dataset characteristics: COCA is centered on a cropped cardiac/chest region, which reduces field-of-view, limits anatomical variability, and provides fewer training slices in our split. In contrast, AAPM offers broader anatomical coverage, including more complete thoracic and abdominal structures, and a substantially larger training set with approximately $600+$ additional slices. Consequently, models trained on AAPM may learn a stronger and more transferable prior, leading to cases where Training AAPM$\rightarrow$Testing COCA matches or exceeds Training COCA$\rightarrow$Testing COCA in absolute PSNR. We include this clarification to avoid confusion when interpreting cross-domain results.

Another active research direction explores implicit neural representations (INRs) and 3D Gaussian representations (3DGS) for tomographic reconstruction. INRs model volumetric data as continuous coordinate-based neural fields, such as SIREN or NeRF~\cite{sitzmann2020implicit,corona2022mednerf,shen2022nerp,kerbl20233d,li20253dgr,wu2025discretized}, while 3D Gaussian splatting represents scenes with learnable Gaussian primitives. These approaches can reconstruct a subject by directly optimizing the representation from its own measurements, thereby reducing reliance on cross-domain generalization. However, they typically require subject-specific optimization, dense or multi-view acquisitions, and substantial optimization time, which can limit applicability to sparse-view or time-critical clinical CT. In contrast, CDPIR learns a population-level diffusion prior that generalizes across scanners and protocols and embeds it into a physics-guided iterative loop for robust reconstruction. These paradigms are complementary, and future hybrid approaches may combine INR/3DGS parameterizations with diffusion-based priors to unify continuous spatial modeling and cross-domain generalization.

A central advantage of CDPIR is its ability to train across heterogeneous datasets, potentially spanning hundreds of imaging domains. By learning a transferable cross-distribution prior that combines shared structural information with domain-specific modulation, CDPIR adapts to specialized tasks, such as cardiac or extremity imaging, while maintaining broad applicability. Its scalability enables unified checkpoints to generalize across datasets and acquisition regimes without task-specific retraining, positioning CDPIR as a promising foundation model for medical image reconstruction with broad deployment potential. Beyond medical imaging, CDPIR's principles of combining generative priors with explicit data consistency could extend to other AI domains, including complex synthesis tasks in diverse modalities and general-purpose text applications\cite{niu2025medical}.


\section{Conclusion}
In summary, CDPIR integrates transformer-based diffusion denoising with explicit data consistency to deliver high-quality SVCT reconstruction. Patch-based positional embeddings enable localized corrections under fidelity constraints while maintaining global coherence, ensuring both fine detail recovery and structural integrity. \red{The conditional diffusion transformer with classifier-free guidance supports cross-distribution prior learning under multi-domain training, while the unified stochastic interpolant framework enhances adaptability to OOD scenarios.} Extensive experiments on AAPM, XCAT, COCA, and real-world PCCT datasets confirm that CDPIR outperforms SOTA baselines such as DiffusionMBIR and DDS. By uniting local precision with global consistency, CDPIR advances the field of medical image reconstruction and paves the way for scalable, generalizable, and clinically deployable generative reconstruction models.

\bibliographystyle{IEEEtran}
\bibliography{refs}
\end{document}